\documentclass[12pt]{article}
\pdfoutput=1
\usepackage{tikz}
\usepackage{putex}
\usepackage{feyn}
\usepackage[vcentermath]{youngtab}
\usepackage{subfig}
\usepackage{lscape}

\usepackage{graphicx}
\usepackage{epstopdf}
\usepackage{enumerate}
\usepackage{cite}
\usepackage{tensor}
\usepackage{slashed}
\usepackage{amsmath}
\usepackage{amssymb}
\usepackage{mathrsfs}
\usepackage{lgrind}
\usepackage{youngtab}

\usepackage{amsmath,amssymb,braket}

\usepackage{bbm}

\usepackage{hyperref}

\numberwithin{equation}{section}

\newcommand {\be} {\begin {equation}}
\newcommand {\ee} {\end {equation}}

\newcommand {\bes} {\begin {equation*}}
\newcommand {\ees} {\end {equation*}}

\def\CH{{\cal H}}

\newcommand{\beq}{\begin{equation}}
\newcommand{\eeq}{\end{equation}}

\def\be{ \begin{equation} }
\def\ee{ \end{equation} }
\def\la#1{\label{#1}}

\def\Tr{{\operatorname{Tr}}}

\def \Rc {\mathcal{R}}
\def \Nc {\mathcal{N}}
\def \Dc {\mathcal{D}}

\def \al{\alpha}

\def \be {\beta}

\def \ga {\gamma}

\def \la {\lambda}

\def \oh {\frac{1}{2}}
\def \id {\mathbbm{1}}

\def \ooo {O(N_1) \times O(N_2) \times O(N_3)}

\def \beq { \begin{equation}}
\def \eeq {\end{equation}}
\def \pr {\partial}
\def \at {\biggl{\vert}}
\def \ra {\rightarrow}
\def \vr {\varnothing}
\def \ty {\tiny\yng}
\def \oh {\frac{1}{2}}

\DeclareMathOperator*{\Pf}{Pf}
\DeclareMathOperator*{\Det}{Det}

\def \l {\left(}
\def \r {\right)}
\def \ll {\langle}
\def \rr {\rangle}

\begin{document}

\preprint{PUPT-2552}

\institution{PU}{Department of Physics, Princeton University, Princeton, NJ 08544}
\institution{PCTS}{Princeton Center for Theoretical Science, Princeton University, Princeton, NJ 08544}
\institution{HU}{Department of Physics, Harvard University, Cambridge, MA 02138}

\title{Spectra of Eigenstates in Fermionic Tensor Quantum Mechanics
}

\authors{Igor R.~Klebanov\worksat{\PU,\PCTS}, Alexey Milekhin\worksat{\PU}, Fedor Popov\worksat{\PU}, 
Grigory Tarnopolsky\worksat{\HU}}

\abstract{
We study the $O(N_1)\times O(N_2)\times O(N_3)$ symmetric quantum mechanics of 3-index Majorana fermions. When the ranks $N_i$ are all equal, this model has a large $N$
limit which is dominated by the melonic Feynman diagrams. We derive an integral formula which computes the number of 
$SO(N_1)\times SO(N_2)\times SO(N_3)$
invariant states for any set of $N_i$. 
For equal ranks the number of singlets is non-vanishing only when $N$ is even, and it exhibits rapid growth: 
it jumps from $36$ in the $O(4)^3$ model to $595354780$ in the $O(6)^3$ model.
We derive bounds on the values of energy, which show  that they scale at most 
as $N^3$ in the large $N$ limit, in agreement with expectations. We also show that the splitting between the lowest singlet and non-singlet states is of order
$1/N$. For $N_3=1$ the tensor model reduces to $O(N_1)\times O(N_2)$ fermionic matrix quantum mechanics, 
and we find a simple expression for the Hamiltonian in terms of the quadratic Casimir operators of the symmetry group. A similar expression is derived
for the complex matrix model with  $SU(N_1)\times SU(N_2)\times U(1)$ symmetry. Finally, we study the $N_3=2$ case of the tensor model, which gives a 
more intricate complex matrix model whose symmetry is only  $O(N_1)\times O(N_2)\times U(1)$. 
All energies are again integers in appropriate units, and we derive a concise formula for the spectrum. 
The fermionic matrix models we studied possess standard 't Hooft large $N$ limits
where the ground state energies are of order $N^2$, while the energy gaps are of order $1$.
}

\date{\centerline{\it Dedicated to the memory of Joe Polchinski}}

\maketitle

\tableofcontents

\section{Introduction and Summary}

In recent literature there has been considerable interest in the quantum mechanical models where the degrees of freedom are fermionic tensors of rank 3 or higher
\cite{Witten:2016iux, Klebanov:2016xxf}.
These models have solvable large $N$ limits dominated by the so-called melonic diagrams. Such novel large $N$ limits were discovered and developed 
in \cite{Gurau:2009tw,Gurau:2011aq,Gurau:2011xq,Bonzom:2011zz,Tanasa:2011ur,Bonzom:2012hw,Dartois:2013he,Carrozza:2015adg,Gurau:2016lzk}, 
mostly in the context of zero-dimensional tensor models with multiple $U(N)$ or $O(N)$ symmetries (for reviews, see \cite{Gurau:2011xp,Tanasa:2015uhr,Gurau:2016cjo}). 
The quantum mechanical tensor models are richer: they have interesting spectra of energy eigenstates and may
have connections with physical systems like the quantum dots. 
More amibitiously, large $N$ tensor quantum mechanics may provide
a dual description of two-dimensional black holes \cite{Almheiri:2014cka,Maldacena:2016upp,Engelsoy:2016xyb,Jensen:2016pah}, 
in the sense of the gauge/gravity duality  \cite{Maldacena:1997re,Gubser:1998bc,Witten:1998qj}. The original motivation \cite{Witten:2016iux} 
for introducing the tensor quantum mechanics is that  they have a large $N$ limit similar to the one in the 
Sachdev-Ye-Kitaev (SYK) model \cite{Sachdev:1992fk,1999PhRvB..59.5341P, 2000PhRvL..85..840G,Kitaev:2015}, but without the necessity of the disorder. Indeed, as shown explicitly in \cite{Klebanov:2016xxf}, the 2- and 4-point functions in the large $N$ tensor models are governed by 
the same Schwinger-Dyson equations as were derived earlier for the SYK-like models \cite{Kitaev:2015,Polchinski:2016xgd,Maldacena:2016hyu,Jevicki:2016bwu,Gross:2016kjj}. 

At the same time, there are significant differences between the tensor and SYK-like models. 
An early hint was the different scaling of the corrections to the large $N$ limit \cite{Witten:2016iux} (see also the further work
in \cite{Bonzom:2017pqs,Klebanov:2017nlk,Ferrari:2017jgw,Benedetti:2018goh}); more recently, additional evidence for the differences
is emerging in the operator spectra and
Hagedorn transition  \cite{Bulycheva:2017ilt,Choudhury:2017tax,Beccaria:2017aqc}. 
The formal structure of the two types of models  is indeed quite different:
the SYK-like models containing a large number of fermions, $N_{\rm SYK}$, have no continuous
symmetries (although an $O(N_{\rm SYK})$ symmetry appears in the replica formalism), while in the tensor models one typically encounters multiple symmetry groups.
For example, in the Gurau-Witten (GW) model \cite{Witten:2016iux} containing 4 Majorana rank-3 tensors, the symmetry is $O(N)^6$; 
there is evidence \cite{Bonzom:2017pqs,Bulycheva:2017ilt} that this model is the tensor counterpart of a 
4-flavor generalization of
the SYK model introduced in \cite{Gross:2016kjj}. 
A simpler tensor quantum mechanics with a single rank-3 Majorana tensor has $O(N)^3$ symmetry \cite{Klebanov:2016xxf} and is the tensor counterpart of the basic 
SYK model with real fermions.
The quantum mechanics of complex rank-3 fermionic tensor, which has $SU(N)^2 \times O(N)\times U(1)$ symmetry \cite{Klebanov:2016xxf}, is the tensor counterpart 
of the variant of SYK model where real fermions are replaced by complex ones \cite{Sachdev:2015efa}.  

The absence of disorder and the presence of the continuous symmetry groups in the tensor models endows them with a number of theoretical advantages, but also makes
them quite difficult to study. In the tensor models any invariant operator should be meaningful and be assigned a definite scaling dimension in the large $N$
limit. While the simplest scaling dimensions coincide with those in the corresponding SYK-like models, the operator spectrum in tensor models is much richer: the number
of $2k$-particle operators grows as $2^k k!$ \cite{Bulycheva:2017ilt,Choudhury:2017tax,Beccaria:2017aqc}. 

Beyond the operator spectrum, it is interesting to investigate
the spectrum of eigenstates of the Hamiltonian. While this spectrum is discrete and bounded for finite $N$, the low-lying states become dense for large $N$ leading
to the (nearly) conformal behavior where it makes sense to calculate the operator scaling dimensions. In the SYK model, the number of states is $2^{N_{\rm SYK}/2}$, and
numerical calculations of spectra
have been carried out for rather large values of $N_{\rm SYK}$ \cite{Garcia-Garcia:2016mno,Cotler:2016fpe}. 
They reveal a smooth distribution of energy eigenvalues, which is almost symmetric under $E\rightarrow -E$;
it exhibits little sensitivity to the randomly chosen coupling constants
$J_{ijkl}$. Such numerical studies of the SYK model  have revealed various interesting physical phenomena, including the quantum chaos. 

The corresponding studies of spectra in the GW model \cite{Witten:2016iux} and the $O(N)^3$
model \cite{Klebanov:2016xxf} have been carried out in \cite{Krishnan:2016bvg,Klebanov:2017,Krishnan:2017ztz,Chaudhuri:2017vrv,Krishnan:2017txw, Krishnan:2017lra, Krishnan:2018hhu}, 
but in these cases the numerical limitations have been more severe -- the number of states 
grows as $2^{N^3/2}$ in the $O(N)^3$ model and as $2^{2 N^3}$ in the GW model. This is why only the $N=2$ GW model and $N=2, 3$ $O(N)^3$ models
have been studied explicitly
so far.\footnote{In \cite{Krishnan:2018hhu} the exact values of the 140 singlet energies in the $O(2)^6$ GW model were found to square to integers. Due to the discrete
symmetries of the GW model, there are only 5 distinct $E<0$ eigenvalues (the singlet spectrum also contains 50 zero-energy states). For these reasons the singlet
spectrum of the $O(2)^6$ GW model exhibits significant gaps. }    
Furthermore, in the tensor models the states need to be decomposed into various representations of the symmetry groups. 
As a result, the details of the energy spectrum in the $O(N)^3$ tensor model are quite different from those in the corresponding SYK model with $N_{\rm SYK}=N^3$ fermion species.

The goal of this paper is to improve our understanding of energy spectra in the tensor models. We will mostly focus on the simplest tensor model with $O(N)^3$ symmetry \cite{Klebanov:2016xxf} and its generalization to 
$O(N_1)\times O(N_2)\times O(N_3)$, where the Majorana tensor degrees of freedom are $\psi^{abc}$ with $a=1, \ldots, N_1$; $b=1, \ldots, N_2$; $c=1, \ldots, N_3$, and 
anti-commutation relations 
\begin{align}
\{ \psi^{abc}, \psi^{a'b'c'}\} = \delta^{aa'}\delta^{bb'}\delta^{cc'}\, . \label{comrel}
\end{align}
The Hamiltonian is taken to be of the ``tetrahedral" form \cite{Carrozza:2015adg,Klebanov:2016xxf}
\begin{align}
H = \frac{g} {4} \psi^{abc}\psi^{ab'c'} \psi^{a'bc'}\psi^{a'b'c} -  \frac{g } {16} N_1 N_2 N_3 (N_1 - N_2 + N_3)\ , 
\label{Htraceless}
\end{align} 
and we have added a shift to make the spectrum traceless. In section \ref{basicsetup} we discuss some essential features of this model, including its discrete symmetries.
In section \ref{somebounds} we will derive lower bounds on the energy in each representation of $SO(N_1)\times SO(N_2)\times SO(N_3)$. We will show that, in the melonic
large $N$ limit where $g N^{3/2}=J$ is kept constant, the most stringent bounds (\ref{boundcorrection}) scale as
$J N^3$, in agreement with expectations for a system with $N^3$ degrees of freedom. On the other hand, the splitting between lowest states in different representations
is found to be of order
$J/N$. Another derivation of this fact, based on effective action considerations, is presented in section \ref{Energygaps}. While this gap vanishes in the large $N$ limit, we expect
the splitting between states in the same representation to vanish much faster, i.e. as $c^{-N^3}$, where $c$ is a positive constant. Such small singlet sector gaps
are needed to account for the large low-temperature entropy, which is given by the sum over melonic graphs and, therefore, has to be of order $N^3$.

If the global symmetry of the quantum mechanical model is gauged, this simply truncates the spectrum to the $SO(N_1)\times SO(N_2)\times SO(N_3)$ invariant states. 
In section \ref{counting} we derive integral formulae
for the number of singlets as functions of the ranks $N_i$. They lead to the conclusion that the singlets are present only when all $N_i$ are even. 
The absence of singlets when some of $N_i$ are odd can often be related to anomalies, which we discuss in section \ref{anomalies}. 
For the $O(N)^3$ model, the number of singlet states is shown in Table 1; it exhibits rapid growth from $2$ for $N=2$, to $36$ for $N=4$, to $595354780$ for $N=6$. 
Thus, even though the $O(4)^3$ model is out of reach of complete numerical diagonalization because it has $64$ Majorana fermions, in contrast to 
the SYK model with $N_{\rm SYK}=64$, it is far from the nearly conformal large $N$ limit. Indeed, since the spectrum is symmetric 
under $E\rightarrow -E$ \cite{Bulycheva:2017ilt}, the number of distinct singlet eigenvalues with $E<0$ cannot exceed $18$. 
Therefore, there are significant gaps in the singlet spectrum of the
$O(4)^3$ model. On the other hand, the presence of the vast number of singlet states for the $O(6)^3$ model suggests that the low-lying
singlet spectrum should be dense for $N=6$ and higher. For large $N$ the number of singlets grows as $\exp \l N^3 \log 2/2 - 3 N^2 \log N /2 \r $. Since all of these states
must fit in an energy interval of order $N^3$, it is plausible that the gaps between low-lying singlet states vanish as $c^{-N^3}$.     

The $O(N_1)\times O(N_2)\times O(N_3)$ tensor model (\ref{Htraceless}) may be viewed as $N_3$ coupled Majorana $N_1\times N_2$ matrices 
\cite{Ferrari:2017ryl,Azeyanagi:2017mre}. As discussed in section \ref{SOmatrix}, 
for $N_3=1$ we find a one-matrix model with $O(N_1)\times O(N_2)$
symmetry, which is exactly solvable because the Hamiltonian may be written in terms of a quadratic Casimir.   
When we set $N_3=2$ we find a complex $N_1\times N_2$ matrix model with $O(N_1)\times O(N_2)\times U(1)$ symmetry. 
It may be studied numerically for values of $N_1$
and $N_2$ as large as 4 and reveals a spectrum which is integer in units of $g/4$. In section \ref{Exactspectrum} 
we explain why this fermionic matrix model is again exactly solvable and derive a concise expression (\ref{h_final}) for its spectrum.
When both $N_1$ and $N_2$ are even, so that the spectrum contains singlet states, we show that the ground state is a singlet. 
In section \ref{SUmatrix} we apply similar methods to another complex matrix model, which was introduced in
\cite{Anninos:2016klf} and
has $SU(N_1)\times SU(N_2)\times U(1)$. It is the $N_3=1$ case of the complex tensor quantum mechanics with  $SU(N_1)\times SU(N_2)\times O(N_3)\times U(1)$
symmetry  \cite{Klebanov:2016xxf}. We show that the Hamiltonian of this model may be expressed in terms of the symmetry charges.  
The solvable matrix models presented in section \ref{matrixmodel} have standard `t Hooft limits when $N_1=N_2=N$ is sent to infinity while $\lambda=g N$ is held fixed.
Then the low-lying states have energies $\sim \lambda N^2$, while the splittings are of order $\lambda$. So, in contrast to the melonic  large $N$
limit, the energy levels don't become dense in the `t Hooft limit of the matrix models. Nevertheless, these fermionic matrix models
are nice examples of exactly solvable `t Hooft limits.   

\section{The rank-3 tensor model and its symmetries}
\label{basicsetup}

The $O(N_1)\times O(N_2)\times O(N_3)$ tensor model is specified by the action
\beq
S = \int \ dt \l  \frac{i}{2} \psi^{abc} \pr_t \psi^{abc} -H \r\ , 
\label{eq:complex}
\eeq
where $H$ is given in (\ref{Htraceless}).
Sometimes it will be convenient to use capital letters $A,B,\ldots$ to denote the multi-index, i.e. $A=(a,b,c)$.
It is easy to see that the Hamiltonian (\ref{Htraceless}) has a traceless spectrum:
\footnote{One can easily compute $\tr(\psi^{abc}\psi^{ab'c'} \psi^{a'bc'}\psi^{a'b'c})=\frac{1 } {4} N_1 N_2 N_3 (N_1 - N_2 + N_3) $ working with $\psi^{abc}$ as with a set of gamma matrices.}
\begin{equation}
\sum_{i} d_i E_i =0\ , \qquad \sum_i d_i = 2^{[N_1 N_2 N_3/2]}\ ,
\end{equation} 
where $d_i$ is the degeneracy of eigenvalue $E_i$.

We can make some general restrictions on the possible values of the energies. Operators $\psi$ obeying
the anti-commutation relation (\ref{comrel}) may be represented as the Majorana $\gamma$-matrices in $N_1 N_2 N_3$--
dimensional Euclidean space.  They have entries which, in our conventions, are
integers divided by $\sqrt{2}$. 
As a result, the Hamiltonian is an integer matrix times $g/16$. 
It is a well-known mathematical fact that such matrices cannot have rational
eigenvalues. Therefore, in units of $g/16$, the energy eigenvalues have to be either
integer or irrational numbers. The explicit results we will find are in agreement with this.

The discrete symmetries of the theory depend on whether some of the ranks are equal. In a $O(N_1)\times O(N)^2$ theory, $N_1\neq N$, we may
study interchange of the two $O(N)$ groups, which acts as $\psi^{abc}\rightarrow \psi^{acb}$.
The invariant operators can be divided into even or odd under the interchange.  
The Hamiltonian (\ref{Htraceless}) is odd \cite{Bulycheva:2017ilt}, which implies that 
the energy spectrum is symmetric under $E\rightarrow -E$. 

Let us construct the operator which implements the interchange
$\psi^{abc}\rightarrow \psi^{acb}$:
\begin{gather}
P_{23} = 2^{N^2(N_1+1)/2}\prod_{a,b=c} \psi^{abc} \prod_{a,b>c} \left(\frac{\psi^{abc}+\psi^{acb}}{\sqrt{2}}\right).
\end{gather}
This operator has the following properties
\begin{gather}
P_{23}^\dagger P_{23} = 1,\quad P^\dagger_{23}=\pm P_{23}, \quad P_{bc}^\dagger \psi^{abc} P_{bc} = (-1)^{N^2(N_1+1)/2+1} \psi^{acb}.
\end{gather}
Due to the last relation one can check
\begin{align}
P^\dagger_{bc} H P_{bc} &=P^\dagger_{bc}\left( \frac{g}{4} \psi^{abc}\psi^{ab'c'} \psi^{a'bc'}\psi^{a'b'c} -  \frac{g}{16} N_1 N_2 N_3 (N_1 - N_2 + N_3)\right) P_{bc}=  \notag\\ 
&=  \frac{g} {4} \psi^{acb}\psi^{ac'b'} \psi^{a'c'b}\psi^{a'cb'} -  \frac{g } {16} N_1 N_2 N_3 (N_1 - N_2 + N_3)=\notag\\
&= -\frac{g}{4} \psi^{abc}\psi^{ab'c'} \psi^{a'bc'}\psi^{a'b'c} + \frac{g}{16} N_1 N_2 N_3 (N_1 - N_2 + N_3) = - H\,,
\end{align}
where we have renamed the repeated indices in the second line and used the anti-commutation relations \eqref{comrel} in the third line. Let us consider any state that is an eigenvector of the $P_{23}$, it exists because $P_{23}$ is either hermitian or antihermitian
\begin{gather}
P_{23}\ket{\lambda} = \lambda \ket{\lambda}, \quad 1=\braket{\lambda|\lambda} = \braket{\lambda|P_{23}^\dagger P_{23}|\lambda} = |\lambda|^2 \braket{\lambda|\lambda} = |\lambda|^2.
\end{gather} 
The energy of such state is equal to zero. Indeed,
\begin{gather}
E = \braket{\lambda|H|\lambda} = - \braket{\lambda|P^{\dagger}_{bc} H P_{bc}|\lambda} = - |\lambda|^2 \braket{\lambda|H|\lambda} = -E, \quad E = 0
\end{gather}

Let us now discuss the case when all three ranks are equal and we have $O(N)^3$ symmetry. Then the invariant operators form irreducible representations of the
group $S_3$ which interchanges the 3 $O(N)$ groups. The Hamiltonian is in the sign representation of degree $1$: it is invariant under the even permutations and
changes sign under the odd ones. Therefore, the symmetry of the Hamiltonian is the alternating group $A_3$, which is isomorphic to $Z_3$. 

The $SO(N_i)$ symmetry charges are
\begin{equation}
Q_{1}^{aa'}= \frac {i}{2} [\psi^{abc},\psi^{a'bc}]\ , \qquad
Q_{2}^{b b'}= \frac {i}{2} [\psi^{ab c},\psi^{a b' c}]\ , \qquad
Q_{3}^{c c'}= \frac {i}{2} [\psi^{ab c},\psi^{a b c'}]\ .
\end{equation}
In addition,
each $O(N_i)$ group contains $Z_2$ parity symmetries which are axis reflections. Inside $O(N_1)$ there are parity symmetries  $P^{a'}$:
for a given $a'$, $P^{a'}$ sends 
$\psi^{a' b c} \rightarrow - \psi^{a' b c}$ for all $b,c$ and leaves all
$\psi^{a b c}, a \neq a'$ invariant. 
It is not hard to see that the corresponding charges are
\begin{equation}
P^{a'}= 2^{N_2 N_3} \prod_{bc} \psi^{a' b c}
\end{equation}
One can indeed check that
\begin{equation}
\left ( P^{a'}\right )^\dagger \psi^{abc} P^{a'} = (-1)^{\delta_{a, a'}+N_2 N_3}  \psi^{abc}\ .
\end{equation}
Similarly, there are $Z_2$ charges inside $O(N_2)$ and $O(N_3)$. A product of two different parity symmetries within the same $O(N_i)$ group is a $SO(N_i)$ rotation. 
Therefore, it is enough to consider one such $Z_2$ parity symmetry within each group and $O(N_i)\sim SO(N_i)\times Z_2$.

The anti-unitary time reversal symmetry $\mathcal{T}$ is a general feature of systems of Majorana fermions; it commutes with them and, therefore, with the
Hamiltonian \eqref{Htraceless}:
\beq
\mathcal{T}^{-1} \psi_{abc} \mathcal{T} = \psi_{abc}\ .
\eeq
The action of $\mathcal{T}$ on the eigenstates depends on the total number of the Majorana fermions $N_1 N_2 N_3$
and is well-known in the theory of topological insulators and superconductors \cite{Kitaev:2009mg}.  
If the total number of fermions is divisible by 8, the operator $\mathcal{T}$ acts trivially, so the ground state may be non-degenerate. Otherwise 
$\mathcal{T}$ acts non-trivially and one finds that the ground state must be degenerate.

\section{Energy bounds for the $O(N_1)\times O(N_2)\times O(N_3)$ model}
\label{somebounds}

Since the Hilbert space of the model is finite dimensional, it is interesting to put an upper bound on the absolute value of the energy eigenvalues in each representation of
the symmetry group. In this section we address this question in two different ways. We first derive a basic linear relation between the Hamiltonian, a quadratic Casimir operator,
and a square of a Hermitian operator which is positive definite. This gives bounds which are useful for the representations
where the quadratic Casimir of one of the orthogonal groups is near its maximum allowed value. We also find that the bounds are exactly
saturated for $N_3=2$, but are not stringent when equal ranks become large.
Then in section \ref{refined} we derive more refined bounds which are more stringent in the large $N$ limit and give the expected scaling of the ground state energy.
Furthermore, we derive a finite multiplicative factor which corrects the refined bound and allows us to deduce the ground state energy in the large $N$ limit.

\subsection{Basic bounds}

To derive an energy bound we introduce the hermitian tensor
\begin{align}
& A^{bc,b'c'} = \frac {i} {2} [\psi^{abc}, \psi^{ab'c'}]=
i \psi^{abc}\psi^{ab'c'}-i \frac{N_1}{2}\delta^{bb'}\delta^{cc'}\,   \nonumber \\ 
& (A^{bc,b'c'})^{\dag} =-i \psi^{ab'c'}\psi^{abc} +i \frac{N_1}{2}\delta^{bb'}\delta^{cc'} = i\psi^{abc} \psi^{ab'c'}-i\frac{N_1}{2}\delta^{bb'}\delta^{cc'} = 
A^{bc,b'c'}\, . \label{opA}
\end{align}
If we think of $bc$ as a combined index which takes $N_2 N_3$ values, then $A^{bc,b'c'}$ are the generators of the transformations in
$O(N_2 N_3) \supset O(N_2)\times O (N_3)$.
The quadratic Casimir of $O(N_2 N_3) \supset O(N_2)\times O (N_3)$,
\begin{equation}
C_2^{O(N_2 N_3)}= \frac{1}{2} A^{bc,b'c'}A^{bc,b'c'}\ ,
\end{equation}
and the quadratic Casimir of the $O(N_1)$ symmetry,
\begin{equation}
C_{2}^{O(N_1)}=\frac{1}{2} Q_{1}^{a a'}Q_{1}^{aa'}
\end{equation}
are related by
\begin{gather}
C_2^{O(N_2 N_3)} +  C_{2}^{O(N_1)} = \frac{N_1 N_2 N_3}{8}\left(N_1 + N_2 N_3 - 2\right) \ .
\label{sumruleCas}
\end{gather}
Therefore, for the states which appear in the model, we find the upper bound:
\begin{gather}
C_{2}^{O(N_1)} \leq \frac{1}{8}N_1 N_2 N_3\left(N_1 + N_2 N_3 - 2\right) \ .
\label{Casimirbound}
\end{gather}
This bound is saturated only if $C_2^{O(N_2 N_3)}=0$ so that the state is invariant under $SO(N_2 N_3)$.

The Hamiltonian may be written as
\begin{align}
H = -\frac{g} {4} A^{bc,b'c'}A^{bc',b'c} + \frac{g}{16}  N_1 N_2 N_3 (N_2- N_3)\ .
\end{align}
Now we note the inequality
\begin{equation}
C_2^{O(N_2 N_3)} \pm \frac{1}{2} A^{bc,b'c'}A^{bc',b'c}= 
\frac{1}{4}(A^{bc,b'c'}\pm A^{bc',b'c})^2 \geq 0
\end{equation}
Combining this with (\ref{sumruleCas}) we get
\begin{align}
\frac{2} {g} H\;  
\begin{cases} 
\leq  \frac{1}{8}N_1 N_2 N_3\left(N_1+N_2-N_3 + N_2 N_3 - 2\right)- C_{2}^{O(N_1)} \,, \\
\geq  C_{2}^{O(N_1)} -  \frac{1}{8}N_1 N_2 N_3\left(N_1-N_2+ N_3 + N_2 N_3 - 2\right)\,.
\end{cases}
\label{genboundone}
\end{align}
In an analogous fashion we can also derive the bounds in terms of $C_2$:
\begin{align}
\frac{2} {g} H\; \begin{cases}
\leq  \frac{1}{8}N_1 N_2 N_3\left(N_2+N_3-N_1 + N_1 N_3 - 2\right) -
C_{2}^{O(N_2)}\,,\\
\geq C_{2}^{O(N_2)} -  \frac{1}{8}N_1 N_2 N_3\left(N_2-N_3+ N_1 + N_1 N_3 - 2\right)  
\end{cases}
\label{genboundtwo}
\end{align}
and similarly in terms of $C_{2}^{O(N_3)}$.

An interesting special case, which we will consider in section \ref{matrixmodel}, is $N_3=2$ where we find a complex $N_1\times N_2$
matrix model. For the singlet states where $C_{2}^{O(N_1)}=C_{2}^{O(N_2)}=0$ the most stringent bound we get from (\ref{genboundone}) and (\ref{genboundtwo}) is
\begin{align}
|H| \leq \frac{g}{8}  N_1 N_2 (N_1+ N_2)\ .
\label{symbound}
\end{align}
In section \ref{matrixmodel} we will show that these bounds are saturated by the exact solution for even $N_1, N_2$.
For $N_1=N_2=N$ we have a $N\times N$ matrix quantum mechanics which
possesses a 't Hooft large $N$ limit where $gN=\lambda$ is held fixed.
In this limit, the ground state energy is $E_0=-{\lambda\over 4} N^2$, which has the expected scaling with $N$ for a matrix model.

More generally, if at least one of the ranks is even (we will call it $N_3$), we may introduce the operators \cite{Krishnan:2017txw}
\begin{gather}
	\bar{c}_{abk}= \frac{1}{\sqrt 2} \left (\psi^{ab(2k-1)}+ i \psi^{ab (2k)} \right ), \quad  c_{abk} = \frac{1}{\sqrt 2} \left (\psi^{ab(2k-1)}- i \psi^{ab (2k)} \right )\,, \notag\\
	\left\{c_{abk}, c_{a' b' k'} \right\}=\left\{\bar{c}_{abk}, \bar{c}_{a' b' k'} \right\}=0,\quad \left\{\bar{c}_{abk}, c_{a'b' k'}\right\}= \delta_{a a'} \delta_{b b'} \delta_{k k'}\,, 
\label{ladderoperator}
	\end{gather}
where $ a=1, 2, \ldots, N_1,\; b=1, 2\ldots, N_2$ and $ k=1, \ldots, \frac{N_3} {2}$. 
In this basis the $O(N_1)\times O(N_2)\times U(N_3/2)$ symmetry is manifest.
The Hamiltonian becomes  \cite{Krishnan:2017txw}
\begin{gather}
H = \frac{g}{2}\big(\bar{c}_{abk}\bar{c}_{ab'k'}c_{a'bk'}c_{a'b'k}-\bar{c}_{abk}\bar{c}_{a'bk'}c_{ab'k'}c_{a'b'k}\big)
+ \frac{g}{2} (N_2- N_1) Q
+ \frac{g}{16}  N_1 N_2 N_3 (N_2- N_1)
 \, ,
\label{onHamilt}
\end{gather}
where
$Q= \frac{1}{2} [\bar c_{abk}, c_{abk}]$.
The Hamiltonian is invariant under the charge conjugation symmetry which interchanges $c_{abk}$ with $\bar c_{abk}$. 

For any even $N_3$, using the basis (\ref{ladderoperator}) we define  the 
oscillator vacuum state
by the condition $c_{ab k}\ket{0} = 0$. Since this condition is invariant under $O (N_1 N_2)$, so is $\ket{0}$. Furthermore, all the states 
that are created by operators that are $O (N_1 N_2 )$ invariant are also $O (N_1 N_2)$ invariant and have energy $\frac{g}{16}  N_1 N_2 N_3 (N_2- N_1)$. The number of such states is estimated to be the dimension of the maximal representation for the $O(N_3)$ group 
$\dim_{\rm max} \sim (N_1 N_2)^{N_3^2/8}$ (see apendix \ref{apa} for details).
The relation \eqref{sumruleCas} also simplifies the search for the singlets. 
For example, we can first forget about the group nature of the third index in the approach of \cite{Krishnan:2017txw} and impose the vanishing of the Casimir of the third group
afterwards. By studying the charges under $U(1)\in U(N_3/2)$ we find that the singlet states must have $N_1 N_2 N_3/4$ creation operators acting on  $\ket{0}$.

Specifying the bound (\ref{genboundone}) to the equal ranks $N_1=N_2=N_3=N$, we find
\begin{align}
C_{2}^{O(N)_1} -\frac{1}{8} N^{3}(N+2)(N-1)\leq \frac{2} {g} H \leq \frac{1}{8}N^{3}(N+2)(N-1) - C_{2}^{O(N)_1}\,.
\label{genbound}
\end{align}
When the bound (\ref{Casimirbound}) is saturated, the corresponding state must have zero energy. This shows that all  
the states invariant under $O(N^2) \supset O(N)_2\times O(N)_3$ have $E=0$.

For the singlet states (\ref{genbound}) gives
\begin{align}
\frac{4} {g} |H| \leq \frac{1}{4}N^{3}(N+2)(N-1) \,.
\label{singletbound}
\end{align}
For $N=2$, exact diagonalization gives that the ground states is a singlet with energy $E_0=-2g$; this saturates the bound (\ref{singletbound}). 
For $N=3$, exact diagonalization gives a ground state with energy $-\frac{5}{4} \sqrt{41} g\approx - 8.0039 g$, which is in the $(2,2,2)$ representation of $O(3)^3$. 
Since for the $2$ of $SO(3)$, $C_1=3/4$, the bound (\ref{genbound}) is $E_0 \geq - \frac{33}{2} g$. This is satisfied and is far from being saturated. 

In the large $N$ limit, $J= g N^{3/2}$ is held fixed. Thus, we obtain a bound on the lowest singlet energy $E_0$, which is $E_0 \geq -c J N^{7/2}$,
where $c$ is a positive constant. Since we expect the ground state energy to be of order $N^3$, this bound is not very informative at large $N$.
A better bound at large $N$ will be derived in the next section.

\subsection{Refined bounds}
\label{refined}

In this section we present another approach to deriving energy bounds for the $O(N_1)\times O(N_2) \times O(N_3)$ invariant states,  which gives a more stringent bound in the large $N$ limit than the ones in the previous section. 

Consider an arbitrary singlet density matrix $\rho$; this means a density matrix invariant under the 
$O(N_1)\times O(N_2) \times O(N_3)$
rotations. For example, it can be $\rho_s = \ket{s}\bra{s}$, where $\ket{s}$ is an singlet state, or if we have some representation $\mathcal{R}$ of the $O(N_1)\times O(N_2) \times O(N_3)$ with an orthonormal basis $\ket{e_i},i=1..\dim\mathcal{R}$ we can define the projector on this subspace of the Hilbert space
\beq
\rho_\mathcal{R} =\frac{1}{\dim \mathcal{R}} \sum_{i=1}^{\dim \mathcal{R}} \ket{e_i}\bra{e_i},\quad \rho_R = 1, \quad \rho^2_\mathcal{R} = \frac{1}{\dim \mathcal{R}} \rho_\mathcal{R}\,. \label{projector}
\eeq
It is easy to see, that this density matrix is invariant under rotations $O^T \rho_\mathcal{R} O = \rho_\mathcal{R}$ for any $O \in O(N_1)\times O(N_2) \times O(N_3)$.
We can calculate the mean value of the energy for this density matrix as
\beq
E = \tr\left[ \rho_\mathcal{R} H\right] = \frac{g}{4}  \tr \left[\rho \psi^{abc}\psi^{ab'c'} \psi^{a'bc'} \psi^{a'b'c} \right] -  \frac{g } {16} N_1 N_2 N_3 (N_1 - N_2 + N_3)\,.
\eeq
For a fixed $a,b,c$ we can act by the rotation matrices (that act trivially on the singlet density matrix $\rho_s$) and make the interchange $a\leftrightarrow 1,b\leftrightarrow 1,c\leftrightarrow 1$. This argument gives us that
\beq
E =\frac{g}{4}  N_1 N_2 N_3 \tr\left[\rho_\mathcal{R} h\right] -   \frac{g } {16} N_1 N_2 N_3 (N_1 - N_2 + N_3), \quad h=\psi^{111}\psi^{1b'c'} \psi^{a'1c'}\psi^{a'b'1}\,, \label{rotatedenergy}
\eeq
where we sum over the repeated indexes. From now on we consider the density matrix to be of the form \eqref{projector}. Now we can estimate the trace in the formula \eqref{rotatedenergy}. With the use of Cauchy –- Schwarz inequality, we have
\beq
\tr \left[\rho_\mathcal{R} h\right]^2 \leq \tr \left[\rho_\mathcal{R} h^\dagger h\right] = \frac12 \tr \big[\rho_\mathcal{R} \psi^{ab1}\psi^{a1c}\psi^{1bc}\psi^{1b'c'} \psi^{a'1c'}\psi^{a'b'1} \big]\,.
\eeq

Because the density matrix $\rho_\mathcal{R}$ is a singlet we can rotate indices back to get
\beq
\tr \left[\rho_\mathcal{R} h\right]^2 \leq \frac{1}{2 N_1 N_2 N_3} \tr\left[ \rho_\mathcal{R} q^\dagger_{abc} q_{abc}\right], \quad q_{abc} = \psi^{ab'c'}\psi^{a'bc'}\psi^{a'b'c}\,.
\eeq
We can express it is the following way
\beq
\left(\tr \left[\rho_\mathcal{R} h\right] - \frac14\left(N_1 - N_2 + N_3\right)\right)^2  \leq \frac{1}{2 N_1 N_2 N_3} \tr\left[ \rho_\mathcal{R}  q^2_{abc}\right] + \frac{1}{16}\left(N_1 - N_2 + N_3\right)^2\,
\eeq
The square of the operator $q_{abc}$ can be expressed as a sum of Casimir operators due to the virtue of the anticommutation relations.
That gives us the bound on the energies of states in representation $\mathcal{R}$:
\beq
\left|E_\mathcal{R}\right| \leq  \frac{g}{16} N_1 N_2 N_3 \Big(N_1 N_2 N_3 + N_1^2 + N_2^2 + N_3^2 - 4 - \frac{8}{ N_1 N_2 N_3}\sum^3_{i=1} \left(N_i+2\right) C^\mathcal{R}_i\Big)^{1/2}\,, \label{generalbound}
\eeq
where $C^\mathcal{R}_i$ is the value of Casimir operator in the representation $\mathcal{R}$. For the singlet  states this  gives
\beq
\left|E\right| \leq \frac{g }{16}N_1 N_2 N_3(N_1 N_2 N_3 + N_1^2+ N_2^2 + N_3^2 -4)^{1/2}\, .  
\label{nicebound}
\eeq
Since $C_i \geq 0$ this bound applies to all energies.
Let us note that for $N_3=2$ the square root may be taken explicitly:
\begin{gather}
\left|E   \right|_{N_3=2} \leq \frac{g }{8}N_1 N_2 (N_1+N_2)\,,
\label{niceboundtwo}
\end{gather}
which is identical to the earlier result (\ref{symbound}). In section \ref{matrixmodel} we will show that this is saturated when $N_1, N_2$ are even and the ground
state is a singlet.

For the case when $N_1=N_2=N_3=N$ and $N>2$ the bound (\ref{nicebound}) is more stringent than the earlier bound (\ref{singletbound}): 
\begin{gather}
|E| \leq E_{bound}= \frac{g }{16}N^3(N+2) \sqrt{N-1}\,  \label{imprbound}
\end{gather}
In the large $N$ limit, $E_{bound}\rightarrow J N^{3}/16$, which is the expected behavior of the ground state energy; in the melonic limit it scales as $N^3$.
We may expand (\ref{generalbound}) for large  $N$ to find
\begin{equation}
\left|E_\mathcal{R}\right| \leq  \frac{g }{16}N^3(N+2) \sqrt{N-1} \left ( 1- \frac{4}{(N+2)(N-1) N^3} \sum^3_{i=1}  C^\mathcal{R}_i + \ldots \right )\ .
\label{boundcorrection}
\end{equation} 
The discussion of the splittings between non-singlet and singlet states in section \ref{Energygaps} will be in agreeement with the scaling of the second term.

We can try to estimate how close the singlet ground state $\ket{\rm vac}$ comes to the bound \eqref{imprbound} by using the exact propagator $G(t)=\braket{T\psi^{abc}(t)\psi^{def}(0)}$ in the large $N$ limit. To do it let us consider the two states
\beq
\ket{1} = \psi^{111}\ket{\rm vac},\quad
\ket{2} = \partial_t \psi^{111}\ket{\rm vac} \,,
\eeq
where we have introduced $\partial_t \psi_{abc} = i [H,\psi^{abc}]$. We can introduce the angle $\theta$ between these states
\beq
\cos^2\theta = \frac{\left|\braket{1|2}\right|^2}{\braket{1|1}\braket{2|2}} = 2\frac{\left|\bra{\rm vac} \psi^{111}\partial_t \psi^{111}\ket{\rm vac} \right|^2}{\bra{\rm vac} \left(\partial_t \psi^{111}\right)^2 \ket{\rm vac}} = 2\frac{ \left|\bra{\rm vac} \psi^{abc}\partial_t \psi^{abc}\ket{\rm vac}\right|^2}{N_1 N_2 N_3\bra{\rm vac} \left(\partial_t \psi^{abc}\right)^2\ket{\rm vac}} ,\label{theangle}
\eeq
where we have rotated back the indexes back by using the fact that the $\ket{\rm vac}$ is a singlet state. One can notice $H = i\psi^{abc}\partial_t \psi^{abc}$, while $\braket{{\rm vac}|\left(\partial_t \psi^{abc}\right)^2|{\rm vac}}$ is just equal to the bound \eqref{nicebound}, then
\beq
\cos^2\theta =\frac{E_0^2}{E_{bound}^2}, \label{theestimation}
\eeq
where $E_{bound}$ is the bound on the singlet ground state energy \eqref{imprbound}. The other way to estimate this angle $\theta$ can be done in the following way. We shift the Hamiltonian, such that the ground state has the zero energy $\left(H-E_0\right)\ket{\rm vac}= 0$ and calculate the expectation value for the energy for the state $\ket{1}$:
\begin{gather}
\braket{E}_1 =\frac{\braket{1|\left(H-E_0\right)|1}}{\braket{1|1}} = 2 \braket{{\rm vac}|\psi^{111} (H-E_0) \psi^{111}|{\rm vac}} =   2 i \braket{{\rm vac}|\psi^{111}\partial_t \psi^{111}|{\rm vac}},
\end{gather}
at the same time the second moment of the energy is
\begin{gather}
\braket{E^2}_1= \frac{\braket{1|(H-E_0)^2|1}}{\braket{1|1}} = 2 \braket{{\rm vac}|\psi^{111} (H-E_0)^2 \psi^{111}|{\rm vac}} = -2 g^2 \braket{{\rm vac}|\left(\partial_t \psi^{111}\right)^2|{\rm vac}}.
\end{gather}
Where we have used the fact that $\partial_t \psi_{abc} = i [H,\psi^{abc}]$. After that we can notice that \eqref{theangle} can be rewritten as
\begin{gather}
\cos^2\theta= \frac{\braket{E}_1^2}{\braket{E^2}_1}. \label{criteria}
\end{gather}
If $\cos\theta=1$, it means that $\braket{E}_1^2=\braket{E^2}_1$ that can be true only if and if $\psi^{111}\ket{\rm vac}$ is an eigenstate of the Hamiltonian. It would give that the propagator is
\[
G(t) = \braket{\psi^{abc} e^{-i H t}\psi^{a'b'c'}} \propto \delta^{aa'} \delta^{bb'} \delta^{cc'} e^{-i \Delta E |t|}\,.
\]
But as we know the solution for the propagator in the large $N$ limit is a conformal propagator. From this we deduce that the bound can not be saturated. Nevertheless we can estimate the angle $\cos^2\theta$. Indeed, in the large $N$ limit the propagator can be calculated numerically or approximated by a conformal one.
From this we can calculate the $\braket{E}_1$ and $\braket{E^2}_1$. We assume $t>t'=0$, $a=a',b=b',c=c'$ and insert the full basis $\ket{E_n}$ of eigenstates of the Hamiltonian in the propagator to get
\begin{gather}
\braket{\psi_{abc}(t) \psi_{abc}(0)} =\sum_n \left|\braket{{\rm vac}|\psi_{abc}|E_n}\right|^2 e^{-i (E_n-E_0)t} = \int \limits^\infty_0 dE \rho (E) e^{-i E t},\notag\\
\text{where}\quad \rho(E)=\sum_n \left|\braket{{\rm vac}|\psi_{abc}|E_n}\right|^2 \delta(E-E_n+E_0).
\end{gather}
The function $\rho(E)$ is known as a structure factor. From this we can calculate 
\beq
\braket{E}_1 = \int\limits_0^\infty dE\, E \,\rho(E) ,\quad \braket{E^2}_1 = \int\limits^\infty_0 dE\, E^2\,\rho(E), \quad \cos^2\theta = \frac{\braket{E}^2_1}{\braket{E^2}_1}.
\eeq
One can use conformal propagator to estimate this angle, which gives $\cos\theta \approx 0.745$, while 
the numerical calculation  \cite{Sachdev:1992fk} gives $\cos\theta=0.6608$. From this and the formula \eqref{theestimation} we get the ground state energy in the large $N$ limit:
\begin{gather}
E_{0} \rightarrow -\cos\theta\, E_{bound} = -\cos\theta\,\frac{J N^3}{16} \approx -0.041 J N^3\ .
\end{gather}
This answer is close to the numerical result for the ground state energy in the SYK model \cite{Cotler:2016fpe}: $E_0 \approx - 0.04 J N_{\rm SYK}$. 
One can make analogous calculations for the other representations. It gives us in the large $N$ limit the following formula for the gap to the lowest state in a representation $\mathcal{R}$:
\beq
E - E_0=  \frac{J \cos\theta}{4 N^2} \sum^3_{i=1}  C^\mathcal{R}_i \label{energyformula}
\eeq

\section{Sigma model and energy gaps}
\label{Energygaps}

In the large $N$ limit the model \ref{eq:complex} is dominated by melonic diagrams. 
This allows one to write down a closed system of 
Schwinger--Dyson equations for the Green function 
$G^{abc}_{a'b'c'}(t_1-t_2) =  \ll T\psi^{abc}(t_1) \psi^{a'b'c'}(t_2) \rr$ and self-energy $\Sigma^{abc}_{a'b'c'}$ and the bare Green 
function $G^{abc}_{a'b'c',0}(\omega) = i \delta^{a}_{a'}\delta^{b}_{b'}\delta^{c}_{c'}/\omega$
\beq
\begin{split}
&(G^{abc}_{a'b'c'}(\omega))^{-1} = \left(G^{abc}_{a'b'c',0}(\omega)\right)^{-1} - \Sigma^{abc}_{a'b'c'}(\omega)\,, \\
&\Sigma^{abc}_{a' b' c'}(t) =g^2 
G_{a' \beta' \ga'}^{a \beta \gamma}(t) G_{\al' b' \ga'}^{\al b \ga}(t) G_{\al' \be' c'}^{\al \be c}(t)\,. 
\label{eq:SD}
\end{split}
\eeq 
For simplicity we shall introduce the multi-index $A=\left(a,b,c\right)$.  We can look for 
a solution in the diagonal form $G^{AB} = G(t) \delta^{AB}$ and $\Sigma^{AB} = \Sigma(t) \delta^{AB}$.
Then we have the following set of equations:
\beq
\begin{split}
G^{-1}(\omega) = -i \omega - \Sigma(\omega)\,,\quad
\Sigma(t) =J^2 G^3(t)\, .
\label{eq:SD_SYK}
\end{split}
\eeq 
These equations exactly coincide with the Schwinger--Dyson equations of the  SYK model 
and have a conformal solution. 

It was argued in \cite{Choudhury:2017tax} that the system of equations (\ref{eq:SD}) can be obtained from the effective action
\footnote{For clarity, we have omitted the indices in the $G^4$ term. Explicitly, this term reads as 
$G_{a' \beta' \ga'}^{a \beta \gamma} G_{\al' b' \ga'}^{\al b \ga} G_{\al' \be' c'}^{\al \be c} G_{a' b' c'}^{abc}$ }:
\beq
\label{s_eff}
S_\text{eff} = - \log \Pf \l \delta_{AB} \pr_t + \Sigma_{AB} \r + \int dt_1 dt_2  
\l -\Sigma_{AB}(t_1-t_2) G^{AB}(t_2 - t_1) - \frac{g^2}{4} G^4(t_1-t_2) \r
\eeq
This action was recently derived from two-particle irreducible diagrams in \cite{Benedetti:2018goh}.

In the strong coupling limit $J \ra \infty$ one 
can neglect the bare Green function. Then, as first noticed in \cite{Choudhury:2017tax},
the \textit{global} symmetry $O(N)^3$ is promoted to the gauged symmetry $O(N)^3$. Indeed, if we neglect $G_0^{AB}(\omega)$ in
(\ref{eq:SD}) then it is easy to see that we can generate a series of solutions by doing
a gauge transformation:
\beq
\begin{split}
G_{AB}(t_1-t_2)  \ra \l O_{A A'}(t_1) \r^T G_{A' B'}(t_1 -t_2) O_{B B'}(t_2) \\
\Sigma_{AB}(t_1-t_2)  \ra \l O_{A A'}(t_1) \r^T \Sigma_{A' B'}(t_1 -t_2) O_{B B'}(t_2)
\end{split}
\label{eq:subs}
\eeq
where we introduce matrix $O$ in $O(N)^3$ which equals to $O_{AB} = O^1_{\alpha \alpha'} O^2_{\beta\beta'}  O^3_{\gamma\gamma'}$.

The effective action (\ref{s_eff}) is also invariant under these transformations if one omits the term $\pr_t$ in the 
Pfaffian.
For finite $J$, the action ceases to be invariant. If we plug the gauge transformation (\ref{eq:subs}) into
the effective action (\ref{s_eff}), the potential does not change, while we will get a kinetic term for matrices $O^i$ of order $1/J$. Indeed,
%
for the conformal solution we have $\Sigma_{AB} = -\l 1/G\r_{AB}$ and
we can rewrite the kinetic part
of the action as
\beq
-\log \Pf \l \delta_{AB} \pr_t + \Sigma_{AB} \r = -\log \Pf \l \delta_{AB} - \pr_t G_{AB} \r - \log \Pf \l \Sigma_{AB} \r
\eeq
The second term $\log \Pf \l \Sigma_{AB} \r$ is invariant under gauge transformations. Then expanding the Pfaffian in the leading order in derivatives we get
\beq
\label{eq:gt}
\oh \int dt \ \Tr \ \pr_t G_{AB}(t,t') \at_{t' \ra t}
\eeq
Making the gauge transformation (\ref{eq:subs}) of the conformal solution $G_{AB} = \delta_{AB} G$ and plugging into
(\ref{eq:gt}) we get:
\beq
\oh \int dt \ \Tr \l N^3 \pr_t G +  N^2 \sum_{i=1}^3  O^T_i(t) G(t-t') \pr_t O_i(t') +  
\pr_t O^T_i(t) G(t-t') O_i(t') \r \at_{t \ra t'} 
\eeq
Factors $N^2$ come from $\Tr(O_1^T O_1)=N$. Now one has to regularize the limit $t_2 \ra t_1$ but this does not going to affect
$N^2$ factors. The details are worked out in \cite{Yoon:2017nig,Benedetti:2018goh}. The upshot is that $G(t-t') O_i(t')$ becomes $\pr_t O_i(t)/J$ 
up to a normalization constant. This leads to the sigma model action
\beq
\label{eq:SM}
S_{SM} = \frac{\mathcal{A} N^2}{J} \int dt \Tr (  \pr_t O_1^T \pr_t O_1 + \pr_t O_2^T \pr_t O_2 + \pr_t O_3^{T} \pr_t O_3 )\,.
\eeq
The spectrum of such a quantum mechanical sigma model is well-known: the Hamiltonian
is proportional to the quadratic Casimir and the eigenstates are simply representations of $O(N)^3$. In our case:
\beq
H_\text{gauge} = \frac{J}{N^2 \mathcal{A}} \l C_2 \l O_1(N) \r + C_2 \l O_2(N) \r + C_2 \l O_3(N) \r \r \,.
\eeq
We note that this has the same structure as the Casimir correction to the energy bound (\ref{boundcorrection}).
Since for the lowest non-singlet representations $C_2 \sim N$, we find the energy gap between singlets and non-singlets to be of the order $\sim J/N$.

\section{Counting singlet states}

\label{counting}

Suppose we have a free fermionic system of $N$ Majorana fermions $\psi^I$, $I=1,\dots,M$ transforming under some representation $\Rc$ of the gauge group $G$.
We want to compute the number of singlet states in such a system. In order to do it, we use the following method.
The Lagrangian in the Euclidean space reads as:
\beq
L = \psi^I \pr_t \psi^I + \psi^I A_{IJ} \psi^J
\eeq
where $A_{IJ}$ is a real gauge field in the representation $\Rc$. Since Majorana fermions anticommute with each other, $A_{IJ}$ must be anti-symmetric $A_{IJ} = - A_{JI}$. The partition function of the gauged system at the temperature $\beta$ is
\beq
\label{z_gauged}
Z_\text{gauged} = \Nc  \int \Dc \psi\, \Dc A\, \ \exp \l -\int_0^\beta dt \,L \r\,,
\eeq
where we have put the fermionic system on a circle with the circumference $\beta$ with antiperiodic boundary conditions $\psi(t) = - \psi(t+\beta)$. The normalization factor $\Nc$ can be easily recovered if we study the ungauged model. The integration over $\Dc A$ gives the volume of the gauge group and the integral over the fermion variables will yield just the dimension of the Hilbert space because the Hamiltonian of the ungauged theory is equal to zero $H_{ungauge} = 0$. In this case the total number of states is simply $2^{M/2}$:
\beq
\label{z_ungauged}
Z_\text{ungauged} = 2^{M/2} \int \Dc A = \Nc \int \Dc \psi\, \Dc A \ \exp \l -\int_0^\beta dt \ \psi \pr_t \psi \r\,.
\eeq
From now on, we will put $\beta=1$.
If we fix Lorentz gauge $\partial_t A = 0$ with $A$ in the Cartan subalgebra, the Faddeev-Popov determinant gives the Haar measure, while the 
path integral over Majoranna fermions gives the partition function of the system with Hamiltonian $H = -\psi^I A_{IJ} \psi^J$. Therefore the \eqref{z_gauged} can be rewritten as
\beq
Z_\text{gauged} =   \int\, \Dc A\, \Tr\,\left(\exp\left(-\psi^I A_{IJ} \psi^J\right)\right),
\eeq
The expression under the trace is an operator of rotations and can be interpreted as a character of the group acting in the Hilbert space of fermions. By the virtue of the representation theory we know that the integral of the character over a group is equal to the number of the trivial representations, i.e. the number of the singlet states. Therefore,
$ Z_\text{gauged} $ equals the number of singlet states. 
If we insert in \eqref{z_gauged} a Wilson line in some representation $\Rc'$, it gives the character of this representation: 
\beq
\left<\Tr_{\Rc'} \exp \l\oint A\, dt\r \right> = \# \text{states in the representation $\Rc'$.}
\eeq

One can compute the partition function because the integral over $\psi$ in both (\ref{z_gauged}) and (\ref{z_ungauged}) is Gaussian; therefore, the problem boils down to computing the Pfaffian:
\beq
Z_\text{gauged} = 2^{M/2} \int \Dc A \frac{\Pf(\pr_t+A)}{\Pf(\pr_t)}\,.
\eeq
As discussed above, we fix $A$ to be a constant matrix in the Cartan subalgebra. The Faddeev--Popov determinant then yields the normalized Haar measure $d\la^N_G$ on the 
gauge group $G$ \cite{Aharony:2003sx}:
\beq
\int_G d \la^N_G = 1\, .
\eeq
Also, since  $A$ is anti-symmetric, the eigenvalues of $A$ are pairs of 
pure-imaginary numbers $i \la_a, -i \la_a$, $a=1,\dots,\lfloor N/2 \rfloor$.
The ratio of the Pfaffians is
\beq
\label{ratio_pfss}
 \frac{\Pf(\pr_t+A)}{\Pf(\pr_t)}= \prod_{a=1}^{M/2} \cos(\la_a/2)\,.
\eeq
There are different ways to derive this formula. One is to compute the ratio of determinants:
\beq
\label{ratio_dets}
\frac{\Det(\pr_t+A)}{\Det(\pr_t)} =\prod_{a=1}^{M/2} \prod_{n=-\infty}^\infty \frac{ \l 2\pi i \l n+\oh \r+i \la_a \r \l 2\pi i \l n+\oh \r -i \la_a \r}{\left(2 \pi i \l n+\oh \r \right)^2} 
= \prod_{a=1}^{M/2} \cos(\la_a/2)^2\,.
\eeq
After that we note that if we go to the Fourier space, both $\pr_t$ and $A$ are real anti-symmetric matrices, so the ratio of Pfaffians must be a real smooth function of $\la_a$.
Therefore, taking the square root of eq. (\ref{ratio_dets}) we get eq. (\ref{ratio_pfss}). 
Alternatively, we can calculate the Pfaffian of $\pr_t+A$ in  
Fourier space. The result is the following formula:
\beq
\label{final_s}
\# \text{singlet states} =  \int d\la^N_G\, \prod_{a=1}^{M/2} 2 \cos(\la_a/2)\,,
\eeq
where we have got the normalization by studying the ungauged theory \eqref{z_ungauged}.



Let us apply this approach to the case when Majorana fermions live in the fundamental representation of several orthogonal groups.
It is important to distinguish between $SO(2n)$ and $SO(2n+1)$. 
The Cartan subalgebra in the $SO(2n)$ algebra consists of the block diagonal matrices with $2\times 2$ blocks 
\beq
\begin{pmatrix} 
0 & x_i \\ 
-x_i & 0 \\ 
\end{pmatrix}\,,
\eeq
where $x_i$ is a rotation phase ranging from $0$ to $2 \pi$. Geometrically it means that for a fixed $SO(2n)$ transformation, there is a basis in which this transformation
looks like a set of rotations in independent two-planes.
In the $SO(2n+1)$ case the last column/row is zero. It corresponds to a fixed one-dimensional subspace.
\textit{Non-normalized} Haar measure in these two cases reads as:
\begin{gather}
d\la_{SO(2 n)}  = \prod_{i<j}^n \sin \l \frac{x_i-x_j}{2} \r^2 \sin \l \frac{x_i+x_j}{2} \r^2 dx_1 \dots dx_n,\\  
d\la_{SO(2n+1)}  = \prod_{i<j}^n \sin \l \frac{x_i-x_j}{2} \r^2 \sin \l \frac{x_i+x_j}{2} \r^2 \prod_{j=1}^n \sin \l \frac{x_j}{2} \r^2 dx_1 \dots dx_n\,.
\end{gather}

Now we discuss the case where the gauge group is the product of three orthogonal groups $SO(N_1) \times SO(N_2) \times SO(N_3)$, so that the 
gauge field decomposes as 
\beq
A=A^1 \otimes \mathbbm{1} \otimes \mathbbm{1} + \id \otimes A^2 \otimes \id + \id \otimes \id \otimes A^3\,.
\eeq
For even $N_i$ in eq. (\ref{final_s}) eigenvalues $\la_a$ are given by $x_i + y_j + z_k$, $\; -x_i + y_j + z_k$, $\; x_i - y_j + z_k$ and 
$ x_i + y_j - z_k$,
with $i=1,\dots,\lfloor N_1/2 \rfloor$, $j=1,\dots,\lfloor N_2/2 \rfloor$, $k=1,\dots,\lfloor N_3/2 \rfloor$. Variables $x_i,y_j,z_k$ are rotation phases for $SO(N_1)$, $SO(N_2)$ and $SO(N_3)$ respectively. In the case when one of the $N_i$ is odd we have to add a zero eigenvalue to this list. With the use of the equation \eqref{final_s} we can write expicit form of the character of the representation and decompose it in terms of characters of the irreducible representations.
For example, for the $O(2)^3$ model the number of singlets is given by the integral
\beq
\frac{16}{(2\pi)^3}
\int\limits_{-\pi}^\pi dx \int\limits_{-\pi}^\pi dy \int\limits_{-\pi}^\pi dz \cos \left ( \frac {x+y+z}{2} \right ) \cos \left ( \frac {x+y-z}{2} \right ) \cos \left ( \frac {x-y+z}{2} \right ) 
\cos \left ( \frac {-x+y+z}{2} \right )
\ ,\eeq
whose evaluation gives $2$.

For the $O(N)^3$ model the number of singlets for various even $N$ is given in Table \ref{on_table}. 
For odd $N$ it is not hard to see that the integral which gives the number of singlets vanishes; this is related to the fact that each group exhibits
an individual anomaly, which we discuss in the next section.\footnote{
Direct diagonalization of the Hamiltonian for $N=3$ \cite{Klebanov:2017,Krishnan:2017ztz}  reveals that there are no non-degenerate eigenvalues, consistent with this. 
There are 8 ground states with energy $-\frac{5}{4} \sqrt{41} g \approx -8.00391 g$; they transform in the spinorial $(2,2,2)$ representation. 
Substuting the value $C_i=3/4$ into the bound \eqref{generalbound} for the energy gives $-11.53 g$, which 
is quite close to the actual value.}
In the next section \ref{sec:large_n} we will show that the number of
singlets grows as $\exp \l N^3 \log 2/2 - 3 N^2 \log N /2 \r $ for large even $N$.
\begin{table}[ht]
\centering
\begin{tabular}[ht]{c|c}
$N$ & \# singlet states \\
\hline
2 & 2 \\
4 & 36 \\
6 & 595354780 \\
\end{tabular}
\caption{Number of singlet states in the $O(N)^3$ model}
\label{on_table}
\end{table}

Using similar methods, the number of singlets can be calculated in the $O(N)^6$ GW model for low values of $N$, and
the results are presented in Table \ref{gw_table}. The fact that there are 140 states for $N=2$ is in agreement with the direct construction of
singlet states in \cite{Krishnan:2018hhu}.
\begin{table}[ht]
\centering
\begin{tabular}[ht]{c|c}
$N$ & \# singlet states \\
\hline
2 & 140 \\
3 & 63358 \\
4 & 114876653804156708 \\
\end{tabular}
\caption{Number of singlet states in the $O(N)^6$ Gurau--Witten model}
\label{gw_table}
\end{table}

We may similarly calculate the number of singlets for the $O(N_1)\times O(N_2)\times O(N_3)$ models. 
When $N_2=N_3=2$, while $N_1$ is even, there are 2 singlets.
For the cases where $N_3=2$, while $N_1$ and $N_2$ are even, some answers are listed in Table \ref{on12_table}.
\begin{table}[ht]
\centering
\begin{tabular}[ht]{c|c}
$(N_1, N_2)$ & \# singlet states \\
\hline
(4,4) & 4 \\
(6,4) & 4 \\
(6,6) & 4 \\
(8,4) & 6 \\
(8,6) & 8 \\
(8,8) & 18 \\
(10,4) &  6\\
(10,6) &  8\\
(10,8) &  20\\
(10,10) &  24\\
\end{tabular}
\caption{Number of singlet states in the $O(N_1)\times O(N_2) \times O(2) $ model}
\label{on12_table}
\end{table}
We note that the growth of the number of singlets for the $O(N)^2 \times O(2)$ model is much slower than for the $O(N)^3$ model.
For low values of $N$ it is not hard to write down explicit expressions for all the singlet states in the oscillator basis; see appendix \ref{explicitsinglets}.
For example, for the $O(4)^2 \times O(2)$ model we find that the 4 singlet energies are $\pm 16 g$ and $\pm 4g$. 

\subsection{Number of singlets for large $N$}
\renewcommand{\al}{x}
\renewcommand{\be}{y}
\renewcommand{\ga}{z}
\label{sec:large_n}
In this section we will estimate the number of singlets in the $SO(N)^3$ model in the large $N$ limit, assuming $N$ to be odd $N=2M$.
For general $N$, the number of singlets is given by the following integral:
\begin{gather} \text{singlet states} = \frac{1}{V^3} \int_{-\pi}^{\pi} [d\al] [d\be] [d\ga] \prod_{i,j,k=1}^{M} 16 \cos \l \frac{\al_i+\be_j+\ga_k}{2} \r
\cos \l \frac{-\al_i+\be_j+\ga_k}{2} \r \times \notag\\ 
\cos \l \frac{\al_i-\be_j+\ga_k}{2} \r \cos \l \frac{\al_i+\be_j-\ga_k}{2} \r \times \\
\prod^M_{i<j} \sin^2 \l \frac{\al_i - \al_j}{2} \r
\sin^2 \l \frac{\al_i + \al_j}{2} \r \sin^2 \l \frac{\be_i - \be_j}{2} \r \sin^2 \l \frac{\be_i + \be_j}{2} \r \sin^2 \l \frac{\ga_i - \ga_j}{2} \r
\sin^2 \l \frac{\ga_i + \ga_j}{2} \r\notag
\end{gather}
Where $V$ is the volume of $SO(N)$. 
When $N$ is large, cosine functions oscillate very rapidly, so the integral localizes near $\al_i=\be_j=\ga_k=0$. Near this point the integrand is positive, so we can exponentiate it:
\beq
\begin{split}
&\# \text{singlet states} = \int_{-\pi}^{\pi} [d\al] [d\be] [d\ga] \exp \l 4 \sum_{n=1}^{\infty} \sum_{i,j,k=1}^M \frac{(-1)^{n+1}}{n} t^n \cos(n \al_i) \cos(n \be_j) \cos(n \ga_k) \r \times \\
&\prod^M_{i<j} \sin^2 \l \frac{\al_i - \al_j}{2} \r
\sin^2 \l \frac{\al_i + \al_j}{2} \r \sin^2 \l \frac{\be_i - \be_j}{2} \r \sin^2 \l \frac{\be_i + \be_j}{2} \r \sin^2 \l \frac{\ga_i - \ga_j}{2} \r
\sin^2 \l \frac{\ga_i + \ga_j}{2} \r
\end{split}
\eeq
Notice that we have introduced a ``regulator'' $t$ which we have to send to one: $t \ra 1$. Similar integrals count operators in theories with tri-fundamental fields
\cite{Beccaria:2017aqc}. In such cases $t=e^{-1/T}$, where $T$ is the temperature. So we are interested in the infinite temperature limit. This case has been studied in detail in
\cite{Beccaria:2017aqc}. Here we perform a similar analysis. As usual, we will encode the saddle-point configuration of the angles $\al,\be,\ga$ using the density function $\rho(\al)$ (obviously
it is the same function for the three $SO(N)$ groups). Moreover this function is symmetric $\rho(\al)=\rho(-\al)$. 
It would be convenient to work with the normalized density $\int_{-\pi}^\pi \ d \al \rho(\al) = 1$. The effective action now reads as:
\begin{gather}
S[\rho]=\frac12 N^3 \int_{-\pi}^{\pi} d \al d \be d \ga \ \rho(\al) \rho(\be) \rho(\ga) 
\sum_{n=1}^\infty \frac{(-1)^{n+1}t^n}{n} \cos(n \al) \cos(n \be) \cos(n \ga) + \notag\\
+\frac14 N^2 \int_{-\pi}^{\pi} \ d\al d\al' \rho(\al) \rho(\al')
\log \sin \l \frac{\al-\al'}{2} \r^4 + \frac14 N^2 \int_{-\pi}^{\pi} \ d\be d\be'  \rho(\be) \rho(\be') \log \sin \l \frac{\be-\be'}{2} \r^4 + \notag\\
+\frac14 N^2 \int_{-\pi}^{\pi} d\ga d\ga' \  \rho(\ga) \rho(\ga') \log \sin \l \frac{\ga-\ga'}{2} \r^4 
\label{s_eff}
\end{gather}
In the infinite temperature limit the saddle-point density is non-zero only on a small interval $[-\al_0,\al_0]$ where $\al_0 \sim \sqrt{\frac{2}{N}}$. The leading contribution is coming from
the first term and it equals to $\frac12 N^3 \log 2$. But this yields simply the dimensions of the Hilbert space, which is $2^{\frac12 N^3}$. The subleading term is coming from the second term in
(\ref{s_eff}). Fortunately, we will not need the exact value of $\al_0$ because of the logarithmic behaviour:
\begin{gather}
 \int_{-\al_0}^{\al_0} d\al d\al' \  \rho(\al) \rho(\al') \log \sin \l \frac{\al-\al'}{2} \r^4 \sim 4 \int_{-\al_0}^{\al_0} d\al d\al' \  \rho(\al) \rho(\al') \log{(\al-\al')}  \sim \notag\\ 
\sim 4 \int_{-\al_0}^{\al_0} d\al d\al' \  \rho(\al) \rho(\al') \log{\al_0} = 4 \log {\al_0} \sim -  2 \log N
\end{gather}
Therefore the subleading term is $-\frac34 N^2 \log N$. So, in total we have
\beq
\#\text{singlet states} \sim \exp \l \frac{N^3}{2} \log 2 - \frac{3 N^2}{2} \log N + O(N^2) \r 
\eeq

\renewcommand{\al}{\alpha}
\renewcommand{\be}{\beta}
\renewcommand{\ga}{\gamma}

\subsection{Anomalies}
\label{anomalies}

Since we are studying fermions on a compact space $S^1$ there is a potential global anomaly associated with $\pi_1(G)$. And indeed it is well-known that $\pi_1(SO(N))=\mathbb{Z}_2$.
Corresponding ``large'' gauge transformation has a simple description: the gauge transformation matrix is the identity matrix, apart from one $2\times 2$ block 
\beq
\begin{pmatrix}
\cos(2 \pi t) & -\sin(2 \pi t) \\
\sin(2 \pi t) & \cos(2 \pi t) \\
\end{pmatrix}\,.
\label{2pi_rot}
\eeq
It is easy to see that after such transformation \textit{one chosen} rotation phase $x_i$ will be shifted by $2\pi$: $x_i \ra x_i + 2\pi$. It does not matter which $x_i$ to pick up, since an even number of
$2\pi$-rotation blocks gives, in fact, a trivial element in $\pi_1(SO(N))$.
It has been known for some time \cite{Elitzur:1985xj} that a theory of a single Majorana fermion in the fundamental representation of $SO(N)$ is suffering from this $\mathbb{Z}_2$ anomaly.
It is instructive to see it using our machinery. The Pfaffian in this case reads as:
\beq
\prod_{i=1}^{N/2} \cos(x_i/2)
\eeq
Under the shift $x_j \ra x_j + 2 \pi$ it changes sign. Therefore the theory is not invariant under large gauge transformations.
In our case of $\ooo$ group it means that at least two out of three $N_i$ should be even, otherwise we will have an odd number of anomalous multiplets.
Since this anomaly is associated with only one group we will refer to it as "individual anomaly". It is easy to see that this anomaly is always $\mathbb{Z}_2$(in other words, it squares to one), even
if we add more gauge groups.

If the gauge group is a product $SO(2n_1) \times SO(2n_2)$ there is a new anomaly mixing these two groups. For each group in the product, the large gauge transformation 
consists of identical $2\times 2$ blocks:
\beq
\begin{pmatrix}
\cos(\pi t) & -\sin(\pi t) \\
\sin(\pi t) & \cos(\pi t) \\
\end{pmatrix}\,.
\label{pi_rot}
\eeq
Since there are two gauge groups, at $t=1$ overall $-1$ will cancel. Now \textit{all} phases $x_i$ and $y_j$ are shifted by $\pi$: $x_i \ra x_i + \pi,\ y_j \ra y_j + \pi $. 
The Pfaffian reads as:
\beq
\prod_{i=1}^{n_1} \prod_{j=1}^{n_2} \cos \l \frac{x_i+y_j}{2} \r \cos \l \frac{x_i-y_j}{2} \r\,.
\eeq
Under the large gauge transformation the Pfaffian acquires $(-1)^{n_1 n_2}$. This anomaly means that for $G=SO(2n_1) \times SO(2n_2) \times SO(N_3)$, $N_3$ can be 
odd only if the product $N_1 N_2$ is even. We will call this anomaly "mixed anomaly". This anomaly is not always $\mathbb{Z}_2$ as we will see shortly.

We do not find any more anomalies: using the long exact sequence in homotopy groups one can show that the fundamental group of 
$SO(2n_1) \times SO(2n_2)/\mathbb{Z}_2$\footnote{One has to divide by $\mathbb{Z}_2$ because $g_1 \times g_2$ acts on $\psi$ in the same way as $(-g_1)\times(-g_2)$} is equal to 
$\mathbb{Z}_2 \times \mathbb{Z}_2 \times \mathbb{Z}_2$ or $\mathbb{Z}_4 \times \mathbb{Z}_2$ depending on $n_1$ and $n_2$.  Using the above explicit descriptions of the individual anomalies and the mixed anomaly
we see that:
\begin{itemize}
\item If $n_1$ and $n_2$ are both even, then the square of the mixed anomaly gives a trivial gauge transformation. Indeed, for each gauge group the number $N_i$ of 
$2\pi$-rotation blocks (\ref{2pi_rot}) is even. Therefore, this is the case of $\mathbb{Z}_2 \times \mathbb{Z}_2 \times \mathbb{Z}_2$.
\item If only one of $n_i$, say $n_1$, is odd, then the mixed anomaly squares to the individual anomaly of $SO(2n_1)$, since this group will have an odd number of $2\pi$ rotation blocks. Therefore, 
the anomalies form $\mathbb{Z}_4 \times \mathbb{Z}_2$.
\item Finally, when both $n_1$ and $n_2$ are odd, then the mixed anomaly squares to the sum of the individual anomalies. This is again $\mathbb{Z}_4 \times \mathbb{Z}_2$.
\end{itemize}

\section{Solution of some fermionic matrix models}

\label{matrixmodel}

When $N_3=1$ or $N_3=2$ the $O(N_1)\times O(N_2)\times O(N_3)$ symmetric tensor model (\ref{Htraceless}) simplifies and becomes a fermionic $N_1\times N_2$ matrix model.
In this section we discuss the solution of these models. For the $O(N_1)\times O(N_2)$ real matrix model the Hamiltonian may be expressed in terms of the quadratic Casimir operators,
which shows that all the states within the same group representation have the same energy. 
This also applies to the $SU(N_1)\times SU(N_2)\times U(1)$ symmetric complex fermionic matrix model, which was considered in \cite{Anninos:2016klf}, \cite{Tierz:2017nvl} (see also \cite{Anninos:2015eji}), and will be
further discussed in section \ref{SUmatrix}. However, the $O(N_1)\times O(N_2)\times U(1)$ complex fermionic matrix model is more complicated in that there are
energy splittings within the same representation of the symmetry group. Nevertheless, as we show in section \ref{Exactspectrum} this model is solvable.

\subsection{The $O(N_1)\times O(N_2)$ model}
\label{SOmatrix}

Setting $N_3=1$ in the $O(N_1)\times O(N_2)\times O(N_3)$ symmetric tensor model (\ref{Htraceless}) 
we find a real matrix model with $O(N_1)\times O(N_2)$ symmetry:
\begin{align}
H = \frac{g} {4} \psi^{ab}\psi^{ab'} \psi^{a'b}\psi^{a'b'} -  \frac{g } {16} N_1 N_2  (N_1 - N_2 + 1)\ . 
\label{Hmatrix}
\end{align}
Using the $SO(N_1)$ and $SO(N_2)$ charges
\begin{equation}
Q_{1}^{aa'}= \frac {i}{2} [\psi^{ab},\psi^{a'b}]\ , \qquad Q_{2}^{b b'}= \frac {i}{2} [\psi^{ab},\psi^{a b'}]
\end{equation}
the Hamiltonian may be expressed in terms of the quadratic Casimirs:
\begin{align}
H = -\frac{g}{2} C_{2}^{SO(N_2)} + \frac{g } {16} N_1 N_2  (N_2 - 1)= \frac{g}{2} C_{2}^{SO(N_1)} - \frac{g } {16} N_1 N_2  (N_1 - 1)\ . 
\end{align}
This shows that, under the interchange of $N_1$ and $N_2$, $H\rightarrow -H$; therefore, for $N_1=N_2$ the spectrum is symmetric around zero.
The sum of this Casimir operators is fixed:
\begin{gather}
C_{2}^{SO(N_1)}+ C_{2}^{SO(N_2)} = \frac{1}{2} Q_{1}^{a a'}Q_{1}^{aa'} + \frac{1}{2} Q_{2}^{b b'}Q_{2}^{b b'} = \frac{1} {8} N_1 N_2 (N_1+ N_2- 2)\ .
\label{casimirsum}
\end{gather}
This shows that there are no states which are singlets under both $SO(N_1)$ and $SO(N_2)$. The irreducible representations $(r_1, r_2)$ which appear in the spectrum must satisfy
the condition (\ref{casimirsum}). In appendix \ref{realmatrix} we list these representations for a few low values of $N_1$ and $N_2$.
The complete lists of
the energies and degeneracies are shown in Table \ref{spn1n2o1}.

For $O(N)\times O(N)$ with even $N$, we find that the ground state is a singlet under $O(N)_1$ 
and transforms in the $SO(N)_2$ representation whose Young diagram is
a $\frac{N}{2} \times \frac{N} {2}$ square. 
The ground state has energy $E_0=- g N^2 (N-1)/16$, while the first excited state is in the fundamental of
$O(N)_1$ which has quadratic Casimir $N-1$. Therefore, the energy gap 
\begin{equation}
E_1- E_0 = \frac {g}{2} (N-1)
\ .
\end{equation}
In the 't Hooft large $N$ limit, $g\sim 1/N$ and the gap stays finite. Therefore, unlike the SYK and tensor models, the matrix model cannot exhibit quasi-conformal behavior.
\begin{table}[h]
\centering
\begin{tabular}{cccccccc}
\hline
\multicolumn{1}{|c|}{$(N_{1},N_{2})$}         & \multicolumn{1}{c|}{(2,2)} & \multicolumn{1}{c|}{(2,3)} & \multicolumn{1}{c|}{(2,4)} & \multicolumn{1}{c|}{(3,3)} & \multicolumn{1}{c|}{(3,4)}& \multicolumn{1}{c|}{(4,4)}& \multicolumn{1}{c|}{(5,5)}   \\ \hline 
\multicolumn{1}{|c|}{$\frac{4}{g}E_{\textrm{degeneracy}}$}      & \multicolumn{1}{c|}{-1$_2$} & \multicolumn{1}{c|}{-1$_6$} & \multicolumn{1}{c|}{-2$_6$} & \multicolumn{1}{c|}{-3$_8$} & \multicolumn{1}{c|}{-6$_{8}$} & \multicolumn{1}{c|}{-12$_{10}$}& \multicolumn{1}{c|}{-20$_{224}$} \\ 
\multicolumn{1}{|c|}{}      & \multicolumn{1}{c|}{1$_{2}$} & \multicolumn{1}{c|}{3$_2$} & \multicolumn{1}{c|}{0$_{8}$} & \multicolumn{1}{c|}{3$_8$} & \multicolumn{1}{c|}{-2$_{36}$} & \multicolumn{1}{c|}{-6$_{64}$} & \multicolumn{1}{c|}{-10$_{1024}$}   \\ 
\multicolumn{1}{|c|}{}      & \multicolumn{1}{c|}{} & \multicolumn{1}{c|}{} & \multicolumn{1}{c|}{6$_{2}$} & \multicolumn{1}{c|}{} & \multicolumn{1}{c|}{6$_{20}$} & \multicolumn{1}{c|}{-4$_{54}$}  & \multicolumn{1}{c|}{-4$_{800}$}  \\ 
\multicolumn{1}{|c|}{}      & \multicolumn{1}{c|}{} & \multicolumn{1}{c|}{} & \multicolumn{1}{c|}{} & \multicolumn{1}{c|}{} & \multicolumn{1}{c|}{} & \multicolumn{1}{c|}{4$_{54}$}  & \multicolumn{1}{c|}{4$_{800}$}  \\ 
\multicolumn{1}{|c|}{}      & \multicolumn{1}{c|}{} & \multicolumn{1}{c|}{} & \multicolumn{1}{c|}{}& \multicolumn{1}{c|}{}  & \multicolumn{1}{c|}{} & \multicolumn{1}{c|}{6$_{64}$} & \multicolumn{1}{c|}{10$_{1024}$}    \\ 
\multicolumn{1}{|c|}{}      & \multicolumn{1}{c|}{} & \multicolumn{1}{c|}{} & \multicolumn{1}{c|}{} & \multicolumn{1}{c|}{} & \multicolumn{1}{c|}{} & \multicolumn{1}{c|}{12$_{10}$}  & \multicolumn{1}{c|}{20$_{224}$}      \\ \hline
\multicolumn{1}{l}{}            & \multicolumn{1}{l}{}   & \multicolumn{1}{l}{}   & \multicolumn{1}{l}{}   & \multicolumn{1}{l}{}  
\end{tabular}
\caption{Spectra of the $O(N_{1})\times O(N_{2})$ models.}
\label{spn1n2o1}
\end{table}

\subsection{The $SU(N_1)\times SU(N_2)\times U(1)$ model}
\label{SUmatrix}

In \cite{Klebanov:2016xxf} a class of complex tensor quantum mechanical models with $SU(N_1)\times SU(N_2)\times O(N_3)\times U(1)$ symmetry was introduced.
We will use the Hamiltonian
\begin{align}
H = g\bar{\psi}_{abc}\bar{\psi}_{a'b'c}\psi_{ab'c'}  \psi_{a'bc'} +g (N_1- N_2) Q+  \frac{ g} {4} N_1 N_2 N_3 (N_1- N_2) \,,
\end{align}
where $\psi_{abc}$ with $a=1, \ldots, N_1$, $b=1, \ldots, N_2$ and $c=1,\ldots,N_{3}$ are complex fermions 
with anti-commutation relations $\{\bar{\psi}_{abc},\psi_{a'b'c'}\}=\delta_{aa'}\delta_{bb'}\delta_{cc'}$.
The second and third terms were added to the Hamiltonian to make it traceless and invariant under the charge conjugation symmetry, which
interchanges $\psi_{abc}$ and $\bar \psi_{abc}$.
This means it is invariant under $Q\rightarrow -Q$, where $Q$
is the $U(1)$ charge: 
\begin{equation}
Q= \bar \psi_{abc} \psi_{abc} - \frac{1}{2}N_1 N_2 N_3\ .
\end{equation}

If we set $N_3=1$ we obtain a complex matrix model with  $SU(N_1)\times SU(N_2)\times U(1)$ symmetry\footnote{
This Hamiltonian is related to that in section 4 of 
\cite{Anninos:2016klf} by changing the coefficients of the second and third terms.}
\begin{align}
H = g\bar{\psi}_{ab}\bar{\psi}_{a'b'}\psi_{ab'}\psi_{a'b}
+g (N_1- N_2) Q+  \frac{ g}{4}  N_1 N_2 (N_1- N_2)
\,, \label{susuham}
\end{align}
which is the subject of this section. Note that the index contraction in the first term is different from those in (\ref{onHamiltMatrix}); the
$SU(N_1)\times SU(N_2)\times U(1)$ symmetry fixes it uniquely.
This matrix model has some features in common with the $O(N_1)\times O(N_2)$ from the previous section. In both of them the energy is completely fixed by the quadratic Casimir
operators of the symmetry group factors. Also, neither model contains states invariant under the entire symmetry group.

The $SU(N_{i})$  charges with $i=1,2$ are 
\begin{gather} 
Q_{1}^{\alpha} = \bar \psi_{ab} (T_{1}^{\alpha})_{aa'} \psi_{a'b},\quad Q_{2}^{\alpha} = \bar \psi_{ab} (T_{2}^{\alpha})_{bb'} \psi_{ab'}\ , \quad \alpha=1, 2, \ldots, N_{i}^2-1\ ,
\label{SUcharges}
\end{gather}
where
we used the Hermitian $SU(N_i)$ generators $T_{i}^{\alpha},\ i=1,2,\  \alpha=1,\dots,N_{i}^2-1$, normalized in the standard fashion:
\beq
\Tr ( T_{1}^{\alpha} T_{1}^{\beta} )=\Tr ( T_{2}^{\alpha} T_{2}^{\beta} ) = \oh \delta^{\alpha\beta}\, .
\eeq
Using the completeness relation (no sum over $i$):
\beq
\label{useful}
( T_{i}^{\alpha} )_{aa'} ( T_{i}^{\alpha} )_{bb'} = \frac{1}{2}\Big(\delta_{ab'} \delta_{a'b} - \frac{1}{N_{i}} \delta_{aa'} \delta_{bb'} \Big)\,.
\eeq 
we find that the quadratic Casimirs of $SU(N_2)$ and $SU(N_2)$:
\begin{align} 
& C_{2}^{SU(N_1)}= Q_1^{\alpha}  Q_1^{\alpha}  = \frac{1} {2} \bar{\psi}_{ab}\bar{\psi}_{a'b'}\psi_{ab'}\psi_{a'b} + \frac{1} {2}(N_1-N_2)  Q - {1\over 2 N_1} Q^2  + {1\over 8}N_1 N_2 (2 N_1 - N_2)\ , \notag \\
& C_{2}^{SU(N_2)}= Q_2^{\alpha} Q_2^{\alpha} =- \frac{1} {2} \bar{\psi}_{ab}\bar{\psi}_{a'b'}\psi_{ab'}\psi_{a'b} + \frac{1} {2} (N_2-N_1) Q - {1\over 2 N_2} Q^2+ 
{1\over 8}N_1 N_2 (2 N_2 - N_1)\ .
 \label{Casimirrel}
\end{align}
Adding them, we obtain the constraint 
\begin{gather} 
C_{2}^{SU(N_1)}+ C_{2}^{SU(N_2)}=
\frac{N_{1}+N_{2}}{2 N_{1}N_{2}}\left(\frac{(N_1 N_2)^2}{4} - Q^2\right)
\label{Casimirsum}
\ .
\end{gather}
To have the singlets of $SU(N_1)$ and $SU(N_2)$, we need the RHS to vanish. 
This means that there are only two $SU(N_1)\times SU(N_2)$ singlet states: the ones with $Q=\pm \frac{N_1 N_2}{2}$. These are the oscillator vacuum $|0\rangle$, which is annihilated
by all $\psi_{ab}$, and the state $|0'\rangle= \prod_{a,b} \bar{\psi}_{ab} |0\rangle $, which is annihilated by all $\bar{\psi}_{ab}$.

The absence of singlets for other values of $Q$ may be seen explicitly as follows. 
The states with charge $- \frac{N_1 N_2}{2} + m$ have the form
\begin{gather}
\bar \psi_{a_{1}b_{1}} \bar \psi_{a_{2}b_{2}} \ldots \bar \psi_{a_{m}b_{m}} 
|0\rangle\ ,
\label{genlevelm}
\end{gather}
but there is no way to contract the indices of $SU(N_1)$ and of $SU(N_2)$; in contrast to the $O(N)$ case, the tensor $\delta_{a_1 a_2}$ is not available.
If $N_1=N_2= N$ there seems to be a state at level $N$ obtained by contracting (\ref{genlevelm}) with $\epsilon_{a_1 \dots a_N}  \epsilon_{b_1 \dots b_N}$, but
this state vanishes due to the Fermi statistics.  

Using (\ref{Casimirrel}) we can express the Hamiltonian  (\ref{susuham}) in terms of the Casimirs:
\begin{gather} 
H = g\Big(2 C_{2}^{SU(N_1)}+\frac{1}{N_{1}}Q^{2}- \frac{1}{4}N^2_{1}N_{2}\Big)\,.
\end{gather}
Therefore, all the states in the same representation of $SU(N_{1})\times SU(N_{2})\times U(1)$ are degenerate, which makes this 
matrix model very simple. In table \ref{spsusu} we list the spectra of the the Hamiltonian  (\ref{susuham})  for a few different values of $N_{1}$ and $N_{2}$. 
\begin{table}[h!]
\centering
\begin{tabular}{cccccccc}
\hline
\multicolumn{1}{|c|}{$(N_{1},N_{2})$}         & \multicolumn{1}{c|}{(1,2)} & \multicolumn{1}{c|}{(1,3)} & \multicolumn{1}{c|}{(2,2)} & \multicolumn{1}{c|}{(2,3)}   \\ \hline 
\multicolumn{1}{|c|}{$\frac{2}{g}E_{\textrm{degeneracy}}$}      & \multicolumn{1}{c|}{-1$_2$} & \multicolumn{1}{c|}{-1$_6$} & \multicolumn{1}{c|}{-4$_3$} & \multicolumn{1}{c|}{-5$_{12}$}  \\ 
\multicolumn{1}{|c|}{}      & \multicolumn{1}{c|}{1$_{2}$} & \multicolumn{1}{c|}{3$_2$} & \multicolumn{1}{c|}{0$_{10}$} & \multicolumn{1}{c|}{-3$_{16}$}  \\ 
\multicolumn{1}{|c|}{}      & \multicolumn{1}{c|}{} & \multicolumn{1}{c|}{} & \multicolumn{1}{c|}{4$_{3}$} & \multicolumn{1}{c|}{1$_{12}$}  \\ 
\multicolumn{1}{|c|}{}      & \multicolumn{1}{c|}{} & \multicolumn{1}{c|}{} & \multicolumn{1}{c|}{} & \multicolumn{1}{c|}{3$_{20}$}  \\ 
\multicolumn{1}{|c|}{}      & \multicolumn{1}{c|}{} & \multicolumn{1}{c|}{} & \multicolumn{1}{c|}{} & \multicolumn{1}{c|}{9$_{4}$}  \\ \hline 
\end{tabular}
\caption{Spectra of the $SU(N_{1})\times SU(N_{2})\times U(1)$ symmetric matrix models.}
\label{spsusu}
\end{table}


\subsection{The $O(N_1)\times O(N_2)\times U(1) $ model}
\label{Exactspectrum}

Setting $N_3=2$ in the $O(N_1)\times O(N_2)\times O(N_3)$ symmetric tensor model (\ref{Htraceless}) 
we find a complex matrix model with $O(N_1)\times O(N_2)\times U(1)$ symmetry.
This model has some features in common with the $SU(N_1)\times SU(N_2)\times U(1)$ model discussed in the previous section; they possess the same $2^{N_1 N_2}$ dimensional
Hilbert space. However, in the present model the symmetry is broken to $O(N_1)\times O(N_2)\times U(1)$ by the Hamiltonian. 
Although the model is still exactly solvable, it is quite interesting in that the energy is not completely fixed by the quadratic Casimir operators of 
$O(N_1)\times O(N_2)\times U(1)$. Also, as we have seen in section \ref{counting}, for even $N_1$ and $N_2$ the model contains singlet states.

To construct the Hilbert space, we define the operators \cite{Krishnan:2017txw}
\begin{gather}
	\bar{\psi}_{ab} =  \frac{1}{\sqrt 2}\left (\psi^{ab1}+ i \psi^{ab2} \right ), \quad \psi_{ab} =  \frac{1}{\sqrt 2}\left (\psi^{ab1}- i \psi^{ab2} \right )\,, \notag\\
	\{\bar{\psi}_{ab}, \bar{\psi}_{a' b'} \}=\{\psi_{ab}, \psi_{a' b'} \}=0,\quad \{\bar{\psi}_{ab}, \psi_{a'b'}\}= \delta_{a a'} \delta_{b b'}\,,
	\end{gather}
where $a=1, 2, \ldots N_1$ and $b=1, 2\ldots N_2$. 
In this basis, the $O(2)$ charge is
\begin{equation}
\label{u1charge}
\begin{split}
&Q = \frac {1} {2}  [\bar{\psi}_{ab}, \psi_{ab}]= \bar{\psi}_{ab} \psi_{ab} - \frac{1}{2}N_1 N_2\,, \\
&[Q, \bar{\psi}_{ab}] =  \bar{\psi}_{ab}, \quad [Q, \psi_{ab}] = - \psi_{ab}\,,
\end{split}
\end{equation}
while the $SO(N_1)$ and $SO(N_2)$ charges are
\begin{gather}
Q_1^{a a'} = i \left ( \bar{\psi}_{ab} \psi_{a'b} - \bar{\psi}_{a' b} \psi_{a b} \right )\,,\notag\\
Q_2^{b b'} = i \left ( \bar{\psi}_{a b} \psi_{a b'} - \bar{\psi}_{a b'} \psi_{a b} \right )\,.
\end{gather}
Squaring these charges, we find the following expressions for quadratic Casimirs:
\begin{align}
& C_{2}^{O(N_1)}=\frac{1}{2} Q_{1}^{a a'}Q_{1}^{aa'}=  \bar{\psi}_{ab}\bar{\psi}_{ab'}\psi_{a'b}\psi_{a'b'}
+ \bar{\psi}_{ab}\bar{\psi}_{a'b'}\psi_{ab'}\psi_{a'b} + (N_1-1) \left (Q+ \frac {1} {2} N_1 N_2 \right ) \ , \nonumber \\
& C_{2}^{O(N_2)}=\frac{1}{2} Q_{2}^{b b'}Q_{2}^{b b'} =  \bar{\psi}_{ab}\bar{\psi}_{a'b}\psi_{ab'}\psi_{a'b'}- 
\bar{\psi}_{ab}\bar{\psi}_{a'b'}\psi_{ab'}\psi_{a'b}+ (N_2-1) \left (Q+ \frac {1} {2} N_1 N_2 \right )   \ .
\label{quadraticCas}
\end{align}

Setting $k=1$ in (\ref{onHamilt}), we find that
the traceless form of the Hamiltonian is 
\begin{gather}
H = \frac{g}{2}\big(\bar{\psi}_{ab}\bar{\psi}_{ab'}\psi_{a'b}\psi_{a'b'}-\bar{\psi}_{ab}\bar{\psi}_{a'b}\psi_{ab'}\psi_{a'b'}\big) 
+ \frac{g}{2} (N_2- N_1) Q
+ \frac{g}{8}N_1 N_2 (N_2- N_1) \,.
\label{onHamiltMatrix}
\end{gather}
This Hamiltonian exhibits the charge conjugation symmetry which acts as $\bar{\psi}_{ab}\leftrightarrow \psi_{ab}$.  This means that states with opposite eigenvalues
of $Q$ have the same energy.

There is a ``Clifford vacuum" state, which satisfies
\begin{equation}
\psi_{ab} |0\rangle =0 \ , \qquad Q |0\rangle = - \frac{N_1 N_2} {2}|0\rangle\ , \qquad
H |0\rangle = \frac{g}{8}N_1 N_2 (N_2- N_1) |0\rangle\ .
\end{equation}
There is also the conjugate vacuum $|0'\rangle =\prod_{ab} \bar{\psi}_{ab} |0\rangle $ which satisfies
\begin{equation}
\bar{\psi}_{ab} |0'\rangle =0 \ , \qquad Q |0'\rangle = \frac{N_1 N_2} {2}|0'\rangle\ , \qquad
H |0'\rangle = \frac{g}{8}N_1 N_2 (N_2- N_1) |0'\rangle\ .
\end{equation}
Both of these states are invariant not only under $O(N_1)\times O(N_2)$, but under the enhanced symmetry $O(N_1 N_2)$. 
It is interesting to note that the states $|0\rangle$ and  $|0'\rangle$ saturate the energy bound (\ref{generalbound}). Indeed, substituting 
$N_3=2$, $C_2^{O(N_3)}= Q^2= (N_1 N_2)^2/4$, $C_2^{O(N_1)}=C_2^{O(N_2)}=0$ into that equation we find $|E|\leq \frac{g}{8}N_1 N_2 |N_2- N_1|$.
In fact, the bound obtained from (\ref{genboundone}) completely fixes the energy to be $\frac{g}{8} N_1 N_2 (N_2- N_1)$ 
because the states are $O(N_1 N_2)$ invariant and 
$C_2^{O(N_1 N_2)}=0$.

The states with vanishing $O(2)$ charge $Q$ are obtained by acting on $|0\rangle$ with 
$\frac{N_1 N_2} {2}$ creation operators $\bar \psi_{ab}$.
Then, to insure that the state is also a singlet under $SO(N_1)\times SO(N_2)$, we have to contract the indices using the invariant tensors 
$\epsilon_{a_1, \ldots a_{N_1}}$, $\delta_{a_1 a_2}$ and $\epsilon_{b_1, \ldots b_{N_2}}$, $\delta_{b_1 b_2}$. 
Some states invariant under $SO(N_1)\times SO(N_2)\times O(2)$ are listed in Appendix \ref{explicitsinglets}.

\begin{figure}[h!]
                \centering
                \includegraphics[width=16cm]{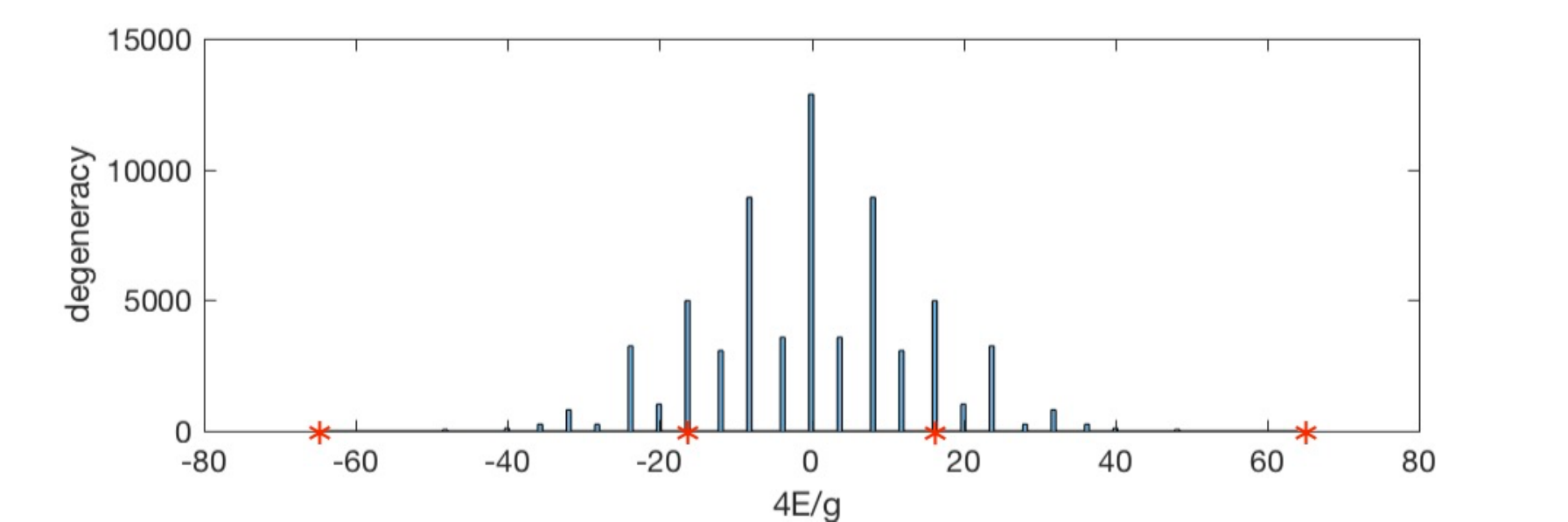}
                \caption{Spectrum of the $O(4)^2\times O(2)$ model. There are four singlet states, and the stars mark their energies.}
                \label{o4o4o2sp}
\end{figure}

For low values of $N_1$ and $N_2$ it is possible to construct the complete spectrum via direct numerical diagonalization. 
If $N_1=N_2$ or if one or both $N_i$ are equal to 2, the spectrum is symmetric under $E\rightarrow -E$ due to the fact that the interchange of two
$O(N)$ groups send $H\rightarrow -H$. For all other values of $N_i$ the spectrum is not symmetric under  $E\rightarrow -E$.
The results for some low values of $N_1, N_2$ are shown in table \ref{spn1n2}. For the $O(4)^2\times O(2)$ model the spectrum is plotted in figure \ref{o4o4o2sp}.

\begin{table}[h]
\centering
\begin{tabular}{ccccccc}
\hline
\multicolumn{1}{|c|}{$(N_{1},N_{2})$}         & \multicolumn{1}{c|}{(2,2)} & \multicolumn{1}{c|}{(2,3)} & \multicolumn{1}{c|}{(3,3)} & \multicolumn{1}{c|}{(2,4)}& \multicolumn{1}{c|}{(4,3)}& \multicolumn{1}{c|}{(4,4)}  \\ \hline 
\multicolumn{1}{|c|}{$\frac{4}{g}E_{\textrm{degeneracy}}$}      & \multicolumn{1}{c|}{-8$_1$} & \multicolumn{1}{c|}{-13$_2$} & \multicolumn{1}{c|}{-20$_6$} & \multicolumn{1}{c|}{-24$_{1}$} & \multicolumn{1}{c|}{-34$_6$}& \multicolumn{1}{c|}{-64$_1$}  \\ 
\multicolumn{1}{|c|}{}      & \multicolumn{1}{c|}{0$_{14}$} & \multicolumn{1}{c|}{-7$_6$} & \multicolumn{1}{c|}{-16$_{18}$} & \multicolumn{1}{c|}{-16$_{2}$} & \multicolumn{1}{c|}{-28$_{24}$} & \multicolumn{1}{c|}{-48$_{55}$}  \\ 
\multicolumn{1}{|c|}{}      & \multicolumn{1}{c|}{8$_1$} & \multicolumn{1}{c|}{-3$_2$} & \multicolumn{1}{c|}{-12$_{16}$} & \multicolumn{1}{c|}{-12$_{16}$} & \multicolumn{1}{c|}{-24$_8$} & \multicolumn{1}{c|}{-40$_{106}$}  \\ 
\multicolumn{1}{|c|}{}      & \multicolumn{1}{c|}{} & \multicolumn{1}{c|}{-1$_{22}$} & \multicolumn{1}{c|}{-8$_{60}$} & \multicolumn{1}{c|}{-8$_{23}$} & \multicolumn{1}{c|}{-22$_{76}$} & \multicolumn{1}{c|}{-36$_{256}$}    \\ 
\multicolumn{1}{|c|}{}      & \multicolumn{1}{c|}{} & \multicolumn{1}{c|}{1$_{22}$} & \multicolumn{1}{c|}{-4$_{42}$} & \multicolumn{1}{c|}{-4$_{16}$} & \multicolumn{1}{c|}{-20$_{40}$} & \multicolumn{1}{c|}{-32$_{810}$}    \\ 
\multicolumn{1}{|c|}{}      & \multicolumn{1}{c|}{} & \multicolumn{1}{c|}{3$_{2}$} & \multicolumn{1}{c|}{0$_{228}$} & \multicolumn{1}{c|}{0$_{140}$} & \multicolumn{1}{c|}{-18$_{14}$} & \multicolumn{1}{c|}{-28$_{256}$}    \\
\multicolumn{1}{|c|}{}      & \multicolumn{1}{c|}{} & \multicolumn{1}{c|}{7$_{6}$} & \multicolumn{1}{c|}{4$_{42}$} & \multicolumn{1}{c|}{4$_{16}$} & \multicolumn{1}{c|}{-16$_{152}$} & \multicolumn{1}{c|}{-24$_{3250}$}   \\ 
\multicolumn{1}{|c|}{}      & \multicolumn{1}{c|}{} & \multicolumn{1}{c|}{13$_2$} & \multicolumn{1}{c|}{8$_{60}$} & \multicolumn{1}{c|}{8$_{23}$} & \multicolumn{1}{c|}{-14$_{168}$} & \multicolumn{1}{c|}{-20$_{1024}$}    \\ 
\multicolumn{1}{|c|}{}      & \multicolumn{1}{c|}{} & \multicolumn{1}{c|}{} & \multicolumn{1}{c|}{12$_{16}$} & \multicolumn{1}{c|}{12$_{16}$} & \multicolumn{1}{c|}{-12$_{40}$} & \multicolumn{1}{c|}{-16$_{4985}$}    \\ 
\multicolumn{1}{|c|}{}      & \multicolumn{1}{c|}{} & \multicolumn{1}{c|}{} & \multicolumn{1}{c|}{16$_{18}$} & \multicolumn{1}{c|}{16$_{2}$} & \multicolumn{1}{c|}{-10$_{170}$}  & \multicolumn{1}{c|}{-12$_{3072}$}  \\ 
\multicolumn{1}{|c|}{}      & \multicolumn{1}{c|}{} & \multicolumn{1}{c|}{} & \multicolumn{1}{c|}{20$_{6}$} & \multicolumn{1}{c|}{24$_{1}$} & \multicolumn{1}{c|}{-8$_{240}$}  & \multicolumn{1}{c|}{-8$_{8932}$}    \\ 
\multicolumn{1}{|c|}{}      & \multicolumn{1}{c|}{} & \multicolumn{1}{c|}{} & \multicolumn{1}{c|}{} & \multicolumn{1}{c|}{} & \multicolumn{1}{c|}{-6$_{194}$}& \multicolumn{1}{c|}{-4$_{3584}$}   \\ 
\multicolumn{1}{|c|}{}      & \multicolumn{1}{c|}{} & \multicolumn{1}{c|}{} & \multicolumn{1}{c|}{} & \multicolumn{1}{c|}{} & \multicolumn{1}{c|}{-4$_{384}$}& \multicolumn{1}{c|}{0$_{12874}$}   \\ 
\multicolumn{1}{|c|}{}      & \multicolumn{1}{c|}{} & \multicolumn{1}{c|}{} & \multicolumn{1}{c|}{} & \multicolumn{1}{c|}{} & \multicolumn{1}{c|}{-2$_{270}$}  & \multicolumn{1}{c|}{4$_{3584}$}  \\ 
\multicolumn{1}{|c|}{}      & \multicolumn{1}{c|}{} & \multicolumn{1}{c|}{} & \multicolumn{1}{c|}{} & \multicolumn{1}{c|}{} & \multicolumn{1}{c|}{0$_{248}$} & \multicolumn{1}{c|}{8$_{8932}$}    \\ 
\multicolumn{1}{|c|}{}      & \multicolumn{1}{c|}{} & \multicolumn{1}{c|}{} & \multicolumn{1}{c|}{} & \multicolumn{1}{c|}{} & \multicolumn{1}{c|}{2$_{640}$}  & \multicolumn{1}{c|}{12$_{3072}$}   \\ 
\multicolumn{1}{|c|}{}      & \multicolumn{1}{c|}{} & \multicolumn{1}{c|}{} & \multicolumn{1}{c|}{} & \multicolumn{1}{c|}{} & \multicolumn{1}{c|}{4$_{384}$} & \multicolumn{1}{c|}{16$_{4985}$}   \\ 
\multicolumn{1}{|c|}{}      & \multicolumn{1}{c|}{} & \multicolumn{1}{c|}{} & \multicolumn{1}{c|}{} & \multicolumn{1}{c|}{} & \multicolumn{1}{c|}{6$_{76}$}  & \multicolumn{1}{c|}{20$_{1024}$}\\ 
\multicolumn{1}{|c|}{}      & \multicolumn{1}{c|}{} & \multicolumn{1}{c|}{} & \multicolumn{1}{c|}{} & \multicolumn{1}{c|}{} & \multicolumn{1}{c|}{8$_{312}$} & \multicolumn{1}{c|}{24$_{3250}$}  \\ 
\multicolumn{1}{|c|}{}      & \multicolumn{1}{c|}{} & \multicolumn{1}{c|}{} & \multicolumn{1}{c|}{} & \multicolumn{1}{c|}{} & \multicolumn{1}{c|}{10$_{216}$} & \multicolumn{1}{c|}{28$_{256}$}  \\ 
\multicolumn{1}{|c|}{}      & \multicolumn{1}{c|}{} & \multicolumn{1}{c|}{} & \multicolumn{1}{c|}{} & \multicolumn{1}{c|}{} & \multicolumn{1}{c|}{14$_{32}$}  & \multicolumn{1}{c|}{32$_{810}$}  \\ 
\multicolumn{1}{|c|}{}      & \multicolumn{1}{c|}{} & \multicolumn{1}{c|}{} & \multicolumn{1}{c|}{} & \multicolumn{1}{c|}{} & \multicolumn{1}{c|}{16$_{128}$} & \multicolumn{1}{c|}{36$_{256}$}     \\ 
\multicolumn{1}{|c|}{}      & \multicolumn{1}{c|}{} & \multicolumn{1}{c|}{} & \multicolumn{1}{c|}{} & \multicolumn{1}{c|}{} & \multicolumn{1}{c|}{18$_{168}$} & \multicolumn{1}{c|}{40$_{106}$}   \\ 
\multicolumn{1}{|c|}{}      & \multicolumn{1}{c|}{} & \multicolumn{1}{c|}{} & \multicolumn{1}{c|}{} & \multicolumn{1}{c|}{} & \multicolumn{1}{c|}{20$_{64}$} & \multicolumn{1}{c|}{48$_{55}$}   \\ 
\multicolumn{1}{|c|}{}      & \multicolumn{1}{c|}{} & \multicolumn{1}{c|}{} & \multicolumn{1}{c|}{} & \multicolumn{1}{c|}{} & \multicolumn{1}{c|}{26$_{10}$} & \multicolumn{1}{c|}{64$_1$}   \\ 
\multicolumn{1}{|c|}{}      & \multicolumn{1}{c|}{} & \multicolumn{1}{c|}{} & \multicolumn{1}{c|}{} & \multicolumn{1}{c|}{} & \multicolumn{1}{c|}{28$_{24}$}& \multicolumn{1}{c|}{}   \\
\multicolumn{1}{|c|}{}      & \multicolumn{1}{c|}{} & \multicolumn{1}{c|}{} & \multicolumn{1}{c|}{} & \multicolumn{1}{c|}{} & \multicolumn{1}{c|}{30$_{6}$}& \multicolumn{1}{c|}{}    \\ 
\multicolumn{1}{|c|}{}      & \multicolumn{1}{c|}{} & \multicolumn{1}{c|}{} & \multicolumn{1}{c|}{} & \multicolumn{1}{c|}{} & \multicolumn{1}{c|}{38$_{2}$} & \multicolumn{1}{c|}{}   \\ \hline
\multicolumn{1}{l}{}            & \multicolumn{1}{l}{}   & \multicolumn{1}{l}{}   & \multicolumn{1}{l}{}   & \multicolumn{1}{l}{}   &\multicolumn{1}{l}{}  
\end{tabular}
\caption{Spectra of the $O(N_{1})\times O(N_{2})\times O(2)$ models, which were obtained by a direct matrix diagonalization of the Hamiltonian
(\ref{onHamilt}) whose spectrum is traceless.
If both $N_1$ and $N_2$ are even, the ground state is non-degenerate and is therefore a singlet.
}
\label{spn1n2}
\end{table}

A remarkable feature of the spectra is that all the eigenvalues of $4H/g$ are integers.  
This suggests that this fermionic matrix model is exactly solvable for any $N_1$ and $N_2$. This is indeed the case, as we now show.
The Hilbert space can be constructed by repeatedly acting with $\bar{\psi}_{ab}$ on the vacuum state $|0\rangle$. One can group the $a,b$ indices into a multi-index $A$, ranging from
1 to $N_1 N_2$. The commutation relations are invariant under the action of $SU(N_1 N_2)$ on the Hilbert space, which preserves the commutation relations.
Let us notice that the first term of Hamiltonian \eqref{onHamilt} is invariant under $SU(N_1)\times O(N_2)\times U(1)$, while the second under 
$O(N_1)\times SU(N_2)\times U(1)$ groups. Therefore, the full Hamiltonian is invariant only under the action of $O(N_1)\times O(N_2)\times$ group. 
The complete Hilbert space is transformed under the $SU(N_1 N_2)$ group that can be split into $SU(N_1) \times SU(N_2)$ representations. In each representation $R$ under $SU(N_2)$, operators $Q_2^{\alpha}$
act by matrices $(T_2^{\alpha})_R$ in the corresponding representation $R$. In turn, these representations can be split into $SO(N_1) \times SO(N_2)$ irreducible 
representations. Since the Hamiltonian has only $SO(N_1) \times SO(N_2)$ symmetry, all the states in such a representation are degenerate (of course, not all the states in
a given $SU(N_1) \times SU(N_2)$ representation are in general degenerate).

Now we take the difference between equations (\ref{quadraticCas}), and also use the difference of equations (\ref{Casimirrel}), to find the following
nice expression for the Hamiltonian: 
\begin{align}
H & = -\frac{g}{2} \Bigg( 2 C_2^{SU(N_1)} - 2 C_2^{SU(N_2)} - C_2^{SO(N_1)} +  C_2^{SO(N_2)}+ \frac{N_{2}-N_{1}}{N_{1}N_{2}} Q^2 
\Bigg) \nonumber \\
& = -\frac{g}{2} \Bigg( 4 C_2^{SU(N_1)}  -  C_2^{SO(N_1)} +  C_2^{SO(N_2)}+ \frac{2}{N_{1}} Q^2 
-\frac{1}{4} N_{1}N_{2} (N_{1}+N_{2})   \Bigg)\ , \label{h_final}
\end{align}
where we used (\ref{Casimirsum}) to obtain the second line from the first.
Due to the $C_2^{SO(N_i)}$ terms, the spectrum is not symmetric under $SU(N_1) \times SU(N_2)$.

Using (\ref{h_final}) we can show that the lowest singlet saturates the energy bound \eqref{niceboundtwo}, i.e. it is a ground state.
For a singlet, $Q$ and the quadratic Casimir operators of $SO(N_1)$ and $SO(N_2)$ vanish.
To minimize the energy we should take a state which has the greatest possible value of $C_2^{SU(N_1)}$ allowed by (\ref{Casimirsum}). 
Thus, it has $C^{SU(N_1)}_2 = \frac{(N_1 + N_2) N_1 N_2}{8}$ and $C^{SU(N_2)}_2 =0$, i.e. it is invariant under $SO(N_1)\times SU(N_2)\times O(2)$. Substituting this into 
(\ref{h_final}) we see that this state has $E=-\frac{g}{8} (N_1 + N_2) N_1 N_2$, i.e. it saturates the bound (\ref{niceboundtwo}).
  This value of Casimir corresponds to the rectangular Young diagram $\lfloor N_1/2 \rfloor \times N_2$ for $SU(N_1)$. 
Similarly, the singlet state with the highest possible energy, $E=\frac{g}{8} (N_1 + N_2) N_1 N_2$, 
has $C^{SU(N_2)}_2 = \frac{(N_1 + N_2) N_1 N_2}{8}$ and $C^{SU(N_1)}_2 =0$, i.e. it is invariant under 
$SU(N_1)\times SO(N_2)\times O(2)$.

To calculate the energies of all states, we need to first decompose the Hilbert space into $SU(N_1)_L \times SU(N_2)_R$ representations and then, in turn, decompose 
these representations into $SO(N_1)_L \times SO(N_2)_R$ representations.  
To find which $SU(N_1)_L \times SU(N_2)_R$ representations $\l L, R \r$ we have in the Hilbert space, we need to compute the following integral over $SU(N_1)_L \times SU(N_2)_R$:
\beq
\text{multiplicity}\l L, R \r = \int \ dU_1 dU_2 \ \exp \l \sum_{n=1}^\infty \frac{(-1)^{n+1}}{n} \Tr U_1^n \ \Tr U_2^n \r \Tr_{L} U_1 \ \Tr_{R} U_2
\eeq
We can always put $U_1$ and $U_2$ in a diagonal form: $U_1 = \diag(w_1,\dots,w_{N_1}), \ U_2 = \diag(q_1,\dots,q_{N_2})$. $w_i$ and $q_i$ are corresponding $SU$ holonomies, i.e. 
$|w_i|=|q_i|=1$ and $w_1 \dots w_{N_1} = q_1 \dots q_{N_2}=1$.

Actually, it is not neccessary to compute the above integral for various representations. It is very well-known that characters of $SU(N_1)$ representations are
Schur polynomials $\Tr_L U_1 = \chi_L(w)$ which form a basis in the space of symmetric functions of $N_1$ variables. This space also contains the so-called power series
polynomials $\Tr U^n_1 = p_n(w) = w_1^n+\dots+w_{N_1}^n$. A conversion from power series $p_n$ to $\chi_L$ can be easily done on a computer. For example,
\begin{gather}
p_1 = \chi_{\ty(1)},\quad 
p_1^2 =  \chi_{\ty(2)} + \chi_{\ty(1,1)}, \notag\\
p_2  =  \chi_{\ty(2)} - \chi_{\ty(1,1)},\quad 
p_1 p_2 = \chi_{\ty(3)} - \chi_{\ty(1,1,1)}\,.
\end{gather}
This suggests the following simple procedure yielding the list of all representations directly. One expands the exponent 
\begin{equation}
\exp \l \sum_{n=1}^\infty \frac{(-1)^{n+1}}{n} x^n\, \Tr\, U_1^n \, \Tr\, U_2^n \r = \exp \l \sum_{n=1}^\infty \frac{(-1)^{n+1}}{n} x^n p_n(w) p_n(q) \r 
\label{}
\end{equation}
in power series in $x$. Then at each level $x^k$ we have a polynomial in $p_l(w)$ and $p_m(q)$. It can be re-expressed in terms of Schur polynomials. This gives the list of representations
under $SU_L(N_1) \times SU_R(N_2)$ at level $k$, i.e. for states where there are $k$ raising operators $\bar \psi$ acting on the vacuum.

After finding the representations under $SU(N_1)_L \times SU(N_2)_R$, we need to decompose then in terms of $SO(N_1)_L \times SO(N_2)_R$ representations.
Recall that both $SU$ and $SO$ representations are classified by Young diagrams. The only difference is that for $SO$ representations one has to subtract
all the traces in each row, where indices are symmetric. 
It means that if we want to extract $SO$ representations from a given $SU$ representation $\la$, we need to consecutivly remove all possible pairs of boxes in each row.
The resulting sequence of Young diagrams give $SO$ representations.

Let us exhibit this method to find the spectrum of the $O(2)^3$ model. We have the following representations under $SU(2)_L \times SU(2)_R$
\footnote{Here we are using the notation $\text{multiplicity}([\dim]_L,[\dim]_R)$
for the $SU(2)_L\times SU(2)_R$ representations and $\text{multiplicity}(\text{spin}_L,\text{spin}_R)$ for $SO(2)_L\times SO(2)_R$ representations. For non-zero spin $J$, the $SO(2)$ representation
is two-dimensional and includes the states with $SO(2)$ charge $Q=\pm J$.}:
\beq
2 ([1],[1]) + 2 ([2],[2]) + ([1],[3]) + ([3],[1])\ .
\eeq
The $[2]$ of $SU(2)$ gives the spin 1 $SO(2)$ representation, whereas the $[3]$ decomposes as $[3] = 2 + 0$.
So we have the following $SO(2)\times SO(2)$ representations:
\beq
2(0,0)+2(1,1)+2(0,0)+(0,2)+(2,0)\ .
\eeq
The two states $(0,0)$ coming from $([1],[3])$
and $([3],[1])$ have energies $\pm 2g$, while all the other states have energy zero. If we label the states by their $O(2)^3$ charges $(Q_1, Q_2, Q_3)$, 
we find, in agreement with \cite{Chaudhuri:2017vrv}, that the states with $E=\pm 2g$ are $(0,0,0)$, while the $14$ zero-energy states are 
\begin{align}
& (1,1,1), (0,0,2), (0,2,0), (2,0,0),(1,1,-1), (1,-1,1), (-1,1,1),\nonumber  \\ 
&  (-1,-1,-1), (0,0,-2), (0,-2,0),  (-2,0,0), (-1,-1, 1),   (-1,1,-1), (1,-1,-1)    
\ .
\end{align}
These states may be decomposed into irreducible representations of the alternating group $A_3$. For example, the state with charges $(1,1,1)$ is invariant
under $A_3$; the 3 states with charges
$ (0,0,2), (0,2,0), (2,0,0) $ can be combined into an invariant combination and a dimension 2 representation; etc. 

As a further check, in appendix \ref{app:spectra}
we calculate the spectrum of the $O(3) \times O(2) \times O(2)$ model using this method.
The results for the energies and their degeneracies agree with the direct diagonalization of the Hamiltonian, whose results are assembled in Table \ref{spn1n2}.
We also note that, due to the charge conjugation symmetry, the energies and
representations at oscillator level $n$ are the same as at level  $N_1 N_2 - n$.

\section*{Acknowledgments}

We are grateful to Ksenia Bulycheva for collaboration at the early stages of this project.
We also thank Dio Anninos, Andrei Bernevig, Sylvain Carrozza, Chethan Krishnan, Juan Maldacena, Kiryl Pakrouski, 
Daniel Roberts, Douglas Stanford and Edward Witten for useful discussions.
The work of IRK was supported in part by the US NSF under Grant No.~PHY-1620059. 
The work of GT was supported in part by  the MURI grant W911NF-14-1-0003 from ARO and by DOE grant de-sc0007870.


\appendix

\section{The eigenvalues of the quadratic Casimir operator}
\label{apa}

In this appendix we describe the value of quadratic Casimir operator for the representations of $O(N)$ and $ SU(N)$ groups in terms of Young diagrams. 
To extract the irreducible representation corresponding to a Young diagram from a generic tensor, we first fill in the boxes with this tensor indices, then we symmetrize over the 
indexes in the rows and after that antisymmetrize the indexes in the columns. 
In the case of the orthogonal group we additionally subtract all possible traces from the tensor.

For the representation of the group $O(N)$ that is described by the Young diagram $Y$ with row lengths $\lambda_i$, the quadratic Casimir operator is equal to
\beq
C^{O(N),Y}_2 = \sum^{\lfloor N/2 \rfloor}_{i=1} \lambda_i \left(\lambda_i + N -2 i\right) \label{apb:casimir}
\eeq
The dimension of this representation reads as:
\beq
\label{apb:dimension}
\text{dim}_\la = \frac{1}{h_\la} \prod_{i=1}^k \frac{(\la_i+N-k-i-1)!}{(N-i)!} \prod_{j=1}^i (\la_i+\la_j+N-i-j)
\eeq
where $h_\la$ is the product of all hook lengths. For each box the hook length is defined as:
\beq
\text{(hook length) = (number of boxes to the right) + (number of boxes below) + 1}
\eeq
The following lemma will be useful for studying the matrix models. Let us consider two groups $O(2n)$ and $O(2m)$ and Young diagram $Y_n$ for group $O(2n)$ such that the 
length of the rows is less then $m$. There is a maximal Young diagram -- a rectangular $ n \times  m$, that we shall denote as $Y_{n\times m}$. We would like to consider a specific 
Young diagram $Y_m =\left(Y_{n\times m}/Y_{n}\right)^T$ for a group $O(2m)$, where T stands for transposition. Then
\beq
C_2^{Y_n}  + C_2^{Y_m} = n^2 m+ n m^2 - n m 
\ .\label{sumofcasimirs}
\eeq
The proof goes as following. Let $\lambda_i$ be the length of rows of the diagram $Y_n$, we introduce $\lambda_0=m,\lambda_{n+1} = 0$. Then
\beq
C_2^{Y_n} = \sum_{i=1}^{n} \lambda_i \left(\lambda_i + 2 (n-i)\right)
\eeq
The value of Casimir operator of $C_2^{Y_m}$ can be expressed as the following. The difference $\lambda_i - \lambda_{i+1}$ is just equal to 
the number of the rows that has length $n-i$. Then
\beq
C_2^{Y_m} = \sum_{i=0}^{n} \left[\left(\lambda_i - \lambda_{i+1}\right) (n-i)^2 +  (n-i) \left(\lambda_i^2-\lambda_{i+1}^2 - \lambda_i + \lambda_{i+1}\right)\right]
\eeq
After that it is easy to see
\begin{gather}
C_2^{Y_m} = m n^2 + n m^2 -n m - \sum_{i=0}^{n} \lambda_i\left(\lambda_i + 2 (n-i) \right)
\end{gather}
So eventually it gives us
\begin{gather}
C_2^{Y_m} + C_2^{Y_n} = m n^2 + n m^2 - n m\ .
\end{gather}
We will call the representation with Young diagram $Y_{n\times m}$ to be maximal and for $O(N)$ group the dimension is
$\dim_{\rm max} \sim n^{m^2/2}$.

We will also need an explicit expression for the quadratic Casimir of $SU(N)$. For a Young diagram $Y$ with row lengths $\la_i$, column lengths $\mu_j$ and total number of boxes $b$
it is given by:
\beq
C^{SU(N),Y}_2 =\frac12 \left (b N + \sum \lambda_i^2 - \sum \mu_j^2 - \frac{b^2}{N}\right )
\ .\eeq

\section{Examples of energy spectra in the matrix models}

\subsection{The $O(N_1)\times O(N_2)$ model for small $N_1, N_2$}
\label{realmatrix}

Let us list the allowed representations for some low values of $N_1$ and $N_2$.
For $O(2)$ we label the representations by the integer charge $Q$ so that the quadratic Casimir $C_2^{O(2)}=Q^2$; 
for $O(3)$ by spin $j$ so that $C_2^{O(3)}=j(j+1)$; for $O(4)\sim SU(2)\times SU(2)$ by
spins $(j_1, j_2)$ so that $C_2^{O(4)}= 2 j_1 (j_1+1)+   2 j_2 (j_2+1)$.

For the $O(2)\times O(2)$ model we find 2 states with $4E/g=-1$ with charges $(\pm 1, 0)$ and 2 states with $4E/g=1$ with charges $(0, \pm 1)$.

For the $O(2)\times O(3)$ model we find 6 states with $4E/g=-1$ which have $SO(3)$ spin 1 and $SO(2)$ charges $\pm 1/2$; and 
2 states with $4E/g=3$ which have $SO(3)$ spin 0 and $SO(2)$ charges $\pm 3/2$.
 
For the $O(3)\times O(3)$ model we find 8 states with $4E/g=-3$ which have spins $(1/2, 3/2)$; and 8 states with $4E/g=3$ which have spins $(3/2, 1/2)$
(note the appearance of half-integral spins which correspond to spinorial representations). 

For the $O(2)\times O(4)$ model we find 6 states with $4E/g=-2$ which have $SO(2)$ charge zero and are in the $SO(4)$ representation $(1,0)+(0,1)$;
8 states with $E=0$ which have $SO(2)$ charges $\pm 1$ and are in the $SO(4)$ representation $(1/2,1/2)$; and
2 states with $4E/g=6$ which have $SO(2)$ charges $\pm 2$ and are $SO(4)$ singlets. 

For the $O(3)\times O(4)$ model we find 8 states with $4E/g=-6$ which have $SO(3)$ spin zero and are in the $SO(4)$ representation $(3/2,0)+(0,3/2)$;
36 states with $4E/g=-2$ which have $SO(3)$ spin $1$ and are in the $SO(4)$ representation $(1/2,1) + (1,1/2)$; and
20 states with $4E/g=6$ which have $SO(3)$ spin $2$ and are in the $SO(4)$ representation  $(1/2,0) + (0,1/2)$.

For the $O(4)\times O(4)$ model we find 10 ground states with $4E/g=-12$ which are $SO(4)_1$ singlets and are in the $SO(4)_2$ representation $(2,0)+(0,2)$;
64 states with $4E/g=-6$ which are in $SO(4)_1$ representation $(1/2, 1/2)$ and in the $SO(4)_2$ representation $(1/2,3/2) + (3/2,1/2)$; etc.

For the $O(6)\times O(6)$ model we find 84 ground states with $4E/g=-45$ which are $SO(6)_1$ singlets and are in the $SO(6)_2$ representation whose Young diagram is
a $3\times 3$ square. The first excited state  has $4E/g=-35$; it transforms as a vector of $SO(6)_1$ and in the representation of $SO(6)_2$ 
whose Young diagram has 3 boxes in the first row, 3 in the second row, and 2 in the third row. 

Due to the relation \eqref{sumofcasimirs} we can state the general correspondence between the representations of $O(N_1)\times O(N_2)$ if $N_1$ and $N_2$ are even. If the state is described by representation $Y_{1}$ for the group $O(N_1)$, then it has the representation $\left(Y_{N_1/2 \times N_2/2}/Y_{1}\right)^T$ for the second group $O(N_2)$.

\subsection{The $O(2)\times O(3) \times U(1)$ model}

\label{app:spectra}

As was described in the main text, first we have to find $SU(2) \times SU(3)$ representations and then decompose into $SO(2) \times SO(3)$ 
irreducible representations. After that we can directly apply the exact formula (\ref{h_final}) for the energy.

Let us list the explicit form of quadratic Casimirs.
For $SO(2)$ the quadratic Casimir is simply $Q^2$, where $Q$ is the charge. 
For $SU(2)$ and $SO(3)$ it equals $j(j+1)$ where $j$ is spin(an integer for $SO(3)$ and half-integer for $SU(2)$).
For $SU(3)$ the quadratic Casimir in our normalization reads as:
\beq
C_2^{SU(3)}(\la)=\oh \l l_1^2+l_2^2-\frac{1}{3}(l_1+l_2)^2+2l_1 \r\,,
\eeq
where $l_1 > l_2 > \dots $ are the row lengths of the Young diagram $\la$ defining the representation $\la$.
For example, $C_2^{SU(3)}(\ty(1))=\frac{4}{3}$, $C_2^{SU(3)}(\ty(2))=\frac{10}{3}$ and $C_2^{SU(3)}(\ty(2,1))=3$  (the last one is the adjoint representation).

The spectrum can be found in Table \ref{sp23}; it coincides with the one in Table \ref{spn1n2}.
\begin{table}
	\centering
\begin{tabular}[h!]{ l l l l  }
  Level & $SU(2) \times SU(3)$ irrep & $SO(2) \times SO(3)$ irrep & $\frac{4}{g}$ Energy \\
\hline
  0 & $\vr \times \vr$ & $\vr \times \vr $  & -3 \\
\hline 
 1 & $\ty(1) \times \ty(1)$ & $\ty(1) \times \ty(1)$ & -1 \\
\hline
  2 & $\ty(2) \times \ty(1)$ & $\ty(2) \times \ty(1)$ & 1 \\
  2 & \phantom{$\ty(2) \times \ty(1)$} & $\vr \times \ty(1)$ & -7 \\
\hline
  2 & $\vr \times \ty(2)$ & $\vr \times \ty(2)$ & 1 \\
  2 & \phantom{$\vr \times \ty(2)$} & $\vr \times \vr$ & 13 \\
\hline
  3 & $\ty(1) \times \ty(2,1) $ & $\ty(1) \times \ty(2)$ & -1 \\
  3 & \phantom{$\ty(1) \times \ty(2,1)$} & $\ty(1) \times \ty(1)$ & 7 \\
\hline
  3 & $\ty(3) \times \vr$ & $\ty(3) \times \vr$ & 3 \\
  3 & \phantom{$\ty(3) \times \vr$} & $\ty(1) \times \vr$ & -13 \\
\hline
  4 & $\vr \times \ty(2)$ & $\vr \times \vr$ & 13 \\
  4 & \phantom{$\vr \times \ty(2)$} & $\vr \times \ty(2)$ & 1 \\
\hline
  4 & $\ty(2) \times \ty(1)$ & $\vr \times \ty(1)$ & -7 \\
  4 & \phantom{$\ty(2) \times \ty(1)$} & $\ty(2) \times \ty(1)$ & 1 \\
\hline
  5 & $\ty(1) \times \ty(1)$ & $\ty(1) \times \ty(1)$ & -1 \\
\hline
  6 & $\vr \times \vr$ & $\vr \times \vr $  & -3 \\
\end{tabular}
\caption{Energy spectrum of the $O(2)\times O(3) \times O(2)$ model. 
Due to the charge conjugation symmetry
for the last $O(2)$ charge, the energies and
representations are invariant under transformation $\text{level} \ra 6-\text{level}$.}
\label{sp23}
\end{table}

\subsection{Explicit form of some singlet states}

\label{explicitsinglets}

The construction of singlet states for the $O(N_1)\times O(N_2)\times O(N_3)$ tensor quantum mechanics is in general a difficult problem, but it simplifies when one of the groups is $O(2)$.
The singlet states, which exist only when $N_1$ and $N_2$ are even, 
may sometimes be written down by inspection in the oscillator basis. In this basis, in addition to the manifest
$SO(N_1)\times SO(N_2)$ symmetry,
 there is manifest discrete $Z_2\times Z_2$ parity symmetry contained inside 
$O(N_1)\times O(N_2)$. 

For example, for the $O(2)^3$ model there are only two singlet states
\begin{gather} 
\epsilon_{a_1 a_2}\delta_{b_1 b_2} \bar{\psi}_{a_1b_1}  \bar{\psi}_{a_2b_2} |0\rangle \ , \quad
 \epsilon_{b_1 b_2}\delta_{a_1 a_2}  \bar{\psi}_{a_1b_1}  \bar{\psi}_{a_2b_2} |0\rangle \ ,
\end{gather}
since due to the Fermi statistics the other two invariant contractions vanish. Under the $Z_2\times Z_2$ symmetry these states are $(-,+)$ and $(+,-)$, respectively.  
In agreement with section \ref{Exactspectrum},
one of these states is invariant under $SU(2)\times SO(2)\times SO(2)$, while the other under  $SO(2)\times SU(2)\times SO(2)$.

Generalizing to any $O(N_1)\times O(2)^2$ model with even $N_1$, we again find only two singlet states. They may be written as 
\begin{gather} 
\epsilon_{a_1, \ldots a_{N_1}}\delta_{b_1 b_2}\ldots  \delta_{b_{N_1-1} b_{N_1}} \bar{\psi}_{a_1b_1} \ldots  \bar{\psi}_{a_{N_1} b_{N_1} } |0\rangle \ , \quad
 \left ( \epsilon_{b_1 b_2}\delta_{a_1 a_2} \bar{\psi}_{a_1b_1}  \bar{\psi}_{a_2b_2} \right )^{N_1/2}|0\rangle  \ .
\end{gather}
One of these states is invariant under $SU(N_1)\times SO(2)\times SO(2)$, while the other under  $SO(N_1)\times SU(2)\times SO(2)$.

For the $O(4)^2\times O(2)$ model there are 4 singlet states
\begin{gather} 
\epsilon_{a_1 a_2 a_3 a_4 } \epsilon_{a_5 a_6 a_7 a_8 } \delta_{b_1 b_5}\ldots  \delta_{b_4 b_8} \bar{\psi}_{a_1b_1} \ldots  \bar{\psi}_{a_8 b_8} |0\rangle \ , \quad
\epsilon_{b_1 b_2 b_3 b_4 } \epsilon_{b_5 b_6 b_7 b_8 } \delta_{a_1 a_5}\ldots  \delta_{a_4 a_8} \bar{\psi}_{a_1b_1}\ldots  \bar{\psi}_{a_8 b_8} |0\rangle \ , \notag \\
\left ( \epsilon_{a_1 a_2 a_3 a_4 } \delta_{b_1 b_2} \delta_{b_3 b_4} \bar{\psi}_{a_1b_1} \ldots  \bar{\psi}_{a_4 b_4} \right )
\left (  \delta_{a_5 a_6} \delta_{a_7 a_8} \delta_{b_5 b_7} \delta_{b_6 b_8} \bar{\psi}_{a_5b_5} \ldots  \bar{\psi}_{a_8 b_8} \right ) |0\rangle \ , \notag \\
\quad
\left ( \epsilon_{b_1 b_2 b_3 b_4 } \delta_{a_1 a_2} \delta_{a_3 a_4} \bar{\psi}_{a_1b_1} \ldots  \bar{\psi}_{a_4 b_4} \right )
\left (  \delta_{b_5 b_6} \delta_{b_7 b_8} \delta_{a_5 a_7} \delta_{a_6 a_8} \bar{\psi}_{a_5b_5} \ldots  \bar{\psi}_{a_8 b_8} \right ) |0\rangle \ .
\end{gather}
The first pair of states have energies $E=\pm 16 g$, saturating the energy bound (\ref{symbound}).
One of these states is invariant under $SU(4)\times O(4)\times O(2)$, while the other under  $O(4)\times SU(4)\times O(2)$.
The second pair of states have energies $E=\pm 4 g$.

Defining the antisymmetric matrix $M_{b_1 b_2}= \bar{\psi}_{a b_1} \bar{\psi}_{a b_2}$, we can write the first two states as
\begin{gather} 
\left ( \tr M^4 \pm \frac {1} {2} (\tr M^2)^2 \right )|0\rangle 
\label{newbasis}
\end{gather}

By analogy with (\ref{newbasis}), for $N$ a multiple of $4$
we may build a set of states by acting on  $|0\rangle$ with traces of powers of $M$. For example, for $N=8$ we can act with $\tr M^{16}$,   $\tr M^2 \tr M^{14}$, etc. 
The number of such terms is $P(8)$, i.e. the number of partitions of $8$ into positive integers, and $P(8)=22$.
For $O(12)^2\times O(2)$ the number of such terms is $P(18)=385$. However, these terms are not linearly independent, so this should be regarded
as an upper bound on the number of invariant states.

More generally, for $O(N)^2\times O(2)$ with $N$ a multiple of $4$, 
this upper bound is $P(N^2/8)$, which grows exponentially with $N$:
\begin{gather} 
P(N^2/8) \rightarrow \frac{2} {N^2 \sqrt 3} \exp \left (\frac {\pi N} {2 \sqrt 3} \right )
\ .
\end{gather}

\bibliographystyle{ssg}
\bibliography{Spectrum}

\end{document}